\documentclass[sigconf]{acmart} 

\copyrightyear{2026}
\acmYear{2026}
\setcopyright{cc}
\setcctype{by}
\acmConference[ISPD '26]{Proceedings of the 2026 International Symposium on Physical Design}{March 15--18, 2026}{Bonn, Germany}
\acmBooktitle{Proceedings of the 2026 International Symposium on Physical Design (ISPD '26), March 15--18, 2026, Bonn, Germany}
\acmPrice{}
\acmDOI{10.1145/3764386.3779611}
\acmISBN{979-8-4007-2314-8/2026/03}

\usepackage{amsmath,amsfonts}
\usepackage{array}
\usepackage{textcomp}
\usepackage{stfloats}
\usepackage{url}
\usepackage{verbatim}
\usepackage{graphicx}
\usepackage{enumitem}
\usepackage{multirow}
\usepackage{tabularx}
\usepackage{listings}
\usepackage{makecell}
\usepackage{wasysym}
\usepackage{siunitx} 
\usepackage{mathtools}
\usepackage{caption}
\usepackage{subcaption}
\usepackage{amsthm}
\usepackage{float}
\usepackage{comment}
\usepackage[ruled,lined,boxed,linesnumbered]{algorithm2e}
\usepackage{balance}
\usepackage{booktabs} 
\usepackage{microtype}
\usepackage{xspace}
\usepackage{bm}
\usepackage{newunicodechar}
\newunicodechar{μ}{\ensuremath{\mu}}
\usepackage{threeparttable}

\begin{document}

\title{Invited: Toward Sustainable and Transparent Benchmarking for Academic Physical Design Research}

\author{Liwen Jiang}
\affiliation{
  \institution{Fudan University}
  \city{Shanghai}
  \country{PRC}
}
\email{lwjiang24@m.fudan.edu.cn}

\author{Andrew B. Kahng}
\affiliation{
  \institution{UC San Diego}
  \city{La Jolla, CA}
  \country{USA}
}
\email{abk@ucsd.edu}

\author{Zhiang Wang*}
\affiliation{
  \institution{Fudan University}
  \city{Shanghai}
  \country{PRC}
}
\email{zhiangwang@fudan.edu.cn}

\author{Zhiyu Zheng}
\affiliation{
  \institution{Fudan University}
  \city{Shanghai}
  \country{PRC}
}
\email{zyzheng24@m.fudan.edu.cn}

\renewcommand{\shortauthors}{Liwen Jiang, Andrew B. Kahng, Zhiang Wang, and Zhiyu Zheng}

\date{November 2025}

\begin{abstract}
This paper presents RosettaStone~2.0, an open benchmark translation and
evaluation framework built on OpenROAD-Research~\cite{openroad_research}.
RosettaStone~2.0 provides complete RTL-to-GDS reference flows for
both conventional 2D designs and Pin-3D-style face-to-face (F2F) hybrid-bonded 3D designs,
enabling rigorous {\em apples-to-apples} comparison across planar and three-dimensional implementation settings.
The framework is integrated within OpenROAD-flow-scripts 
(ORFS)-Research~\cite{orfs_research};
it incorporates continuous integration (CI)-based regression  testing and 
provides a standardized evaluation pipeline based on the METRICS2.1 convention, 
with structured logs and reports generated by ORFS-Research.
To support transparent and reproducible research, 
RosettaStone~2.0 further provides a community-facing leaderboard,
which is governed by verified pull requests and enforced through Developer Certificate of Origin (DCO) compliance.
\end{abstract}



\maketitle

\section{Introduction}
\label{sec:introduction}

The advancement of 2D and 3D VLSI physical design (PD) research is inherently dependent on 
rigorous benchmarking frameworks built upon complete flows, realistic benchmarks, 
and reproducible methodologies. 
Academic contests have historically provided benchmarks that drive innovation in physical design~\cite{ispd2005, iccad2015}.
However, these benchmarks usually lack the full set of inputs required by modern RTL-to-GDS flows; 
even Liberty, technology LEFs, parasitic models, and
signoff constraints are often missing or only partially specified.
Moreover, different publications may use different libraries, timing
corners, or routing rules under the same testcase name, and scripts are
frequently tied to specific tools or versions. 
As a result, reported quality-of-results (QoR) 
and runtime improvements are difficult to compare in an {\em apples-to-apples} way, especially 
across independent works.

Recent perspectives argue that the next physical design roadmap inflection is driven by
fine-grain heterogeneous 3D integration~\cite{zhao_analytical_2025}, where bonding and backside processing
make it practical to stack multiple active tiers and tailor per-tier device and
interconnect choices to the function being implemented~\cite{brunion_cmos_2025}. 
Despite strong interest in fine-grain 3D integration, there is still no widely 
adopted open-source RTL-to-GDS reference flow for Pin-3D-style face-to-face (F2F)
hybrid bonding \cite{pentapati_pin-3d_2020, pentapati_pin-3d_2024}, 
with maintainable 3D enablements and reproducible stage-wise checkpoints. 
Hence, the integration, fair comparison, and reproducibility of emerging 
3D physical design algorithms across different tool versions and Process Design Kit (PDK)
variants remain challenging.
Reproducible evaluation must increasingly encompass both 
planar and 3D settings, while enabling more rigorous control not only of algorithmic 
choices, but of tool enablements and measurement protocols as well.

Several 3D integration styles have been explored. Through-silicon via (TSV)
based stacking is well-established in commercial products such as high-bandwidth 
memory, but TSVs with 10~$\mu$m-scale pitch consume area and add 
large parasitics~\cite{fischer_wire-bonded_2011, vartanian_cost_2014}.
Monolithic 3D (M3D) uses nanoscale monolithic inter-tier vias and 
sequential low-temperature processing~\cite{jadhav_architecture_2021} to achieve
very high integration density, but remains costly and less mature~\cite{zhao_toward_2025}.
By contrast, face-to-face hybrid bonding bonds two pre-fabricated dies
using hybrid bonding terminals (HBTs) in the back-end-of-line (BEOL), with an
approximately 1~$\mu$m-scale pitch and strong manufacturing 
support~\cite{liu_routing-aware_2024, netzband_05_2023}.
Recent products such as AMD's Zen~4 and Meta's AR SoC adopt F2F hybrid bonding~\cite{munger_zen_2023,wu_3d_2024},
which makes F2F-based 3D integration a realistic and attractive target for
academic 3D PD flows.

Recent academic research works have started to address challenges of
reproducibility and standardization for 3D physical design. 
Open-3DBench~\cite{shi_open3dbench_2025} provides an open 3D-IC backend 
benchmarking framework built on OpenROAD-flow-scripts \cite{ORFS}, covering 
partitioning, placement, routing, extraction, and thermal analysis across 3D flows. 
Meanwhile, the IEEE CEDA DATC Robust Design Flow (RDF)~\cite{chhabria_invited_rdf_2025} 
and the ML EDA Commons~\cite{chhabria_invited_2025} seek to advance shared 
benchmarks, metrics, and evaluation infrastructures based on OpenROAD \cite{OpenROAD}
and ORFS, with a strong focus on artifact evaluation and sustainable research backplanes. 
Despite such efforts, a durable and open-source RTL-to-GDS reference flow for 
Pin-3D-style F2F hybrid bonding -- with maintainable enablements and 
standardized evaluation checkpoints -- remains unavailable. 
To advance the availability of fair and reproducible assessments for 3D PD flows,
we make the following contributions.


%
%

\begin{itemize}[noitemsep, topsep=0pt, leftmargin=*]
\item \textbf{Sustainable benchmarking backplane.}
We establish a maintained infrastructure within OpenROAD- and ORFS-Research,
featuring co-versioned artifacts, CI-based regression, METRICS2.1-compatible
reporting, and community-driven governance.

\item \textbf{Pin-3D reference flow.}
We release an end-to-end F2F 3D RTL-to-GDS reference flow, complete with 3D enablements
(e.g., HBT modeling, tier libraries) and standardized checkpoints for reproducible 
evaluation.

\item \textbf{RosettaStone~2.0 roadmap.}
We outline the extension of this backplane to support systematic benchmark 
translation, synthetic netlist generation, and unified 2D and 3D validation under 
explicit evaluation contracts.
\end{itemize}

In the following, Section~\ref{sec:sustainable_backplane} describes prospects
for OpenROAD-Research and ORFS-Research, and introduces
RosettaStone 2.0 as a sustainable benchmarking backplane supporting 
reproducible  2D and Pin-3D physical design evaluation.
Section~\ref{sec:pin3d_flow} presents our initial Pin-3D enablement and 
RTL-to-GDS reference flow for F2F hybrid bonding.
Section~\ref{sec:pin3d_validation} reports example validation results and 
sensitivity analyses under a consistent evaluation contract.
Section~\ref{sec:rs2_roadmap} outlines the roadmap toward the complete 
RosettaStone~2.0 framework, and we conclude in Section~\ref{sec:conclusion}.

\section{Sustainable Benchmarking Backplane Based on OpenROAD-Research and ORFS-Research}
\label{sec:sustainable_backplane}

OpenROAD-Research~\cite{openroad_research} and ORFS-Research~\cite{orfs_research} 
are designed as open and collaborative
platforms to guide future efforts that will strengthen open EDA infrastructure and 
community collaboration. They preserve algorithmic diversity by hosting
academically validated but potentially non-production code, thus lowering the barrier
to experimentation in emerging areas such as ML-assisted optimization and
agentic flows; supporting education through reproducible implementations; and
fostering global collaboration via contrib-style repositories. In contrast to 
the mainline repositories \cite{OpenROAD} \cite{ORFS}, which focus on 
industrial-grade stability, the research backplanes prioritize transparency, 
reproducibility, and extensibility, making them a natural substrate for 
sustainable benchmarking.

Building on this backplane, we release the Pin-3D RTL-to-GDS reference flow 
as a set of versioned sub-tools within ORFS-Research. The Pin-3D flow provides an
open, reproducible instantiation of fine-grained 3D physical design for
face-to-face hybrid bonding, including curated flow scripts, maintainable
3D enablements, and METRICS2.1-compatible structured logging. 
The structured evaluation kernel enables consistent measurement of
wirelength, routing congestion, total negative slack, design-rule violations,
and runtime across all stages, allowing direct end-to-end comparison between
alternative placement engines and the native OpenROAD-Research flow under
identical downstream conditions.

While OpenROAD-Research and ORFS-Research provide the execution substrate,
systematic benchmarking further requires robust mechanisms for benchmark
translation, normalization, and coverage expansion. To this end, RosettaStone~2.0 
provides an extended backplane that enables systematic translation of legacy 
academic benchmarks, integration of synthetic benchmark generation, and unified 
validation of both 2D and 3D flows under explicit evaluation contracts. 
These contracts define stage boundaries, exported artifacts, and measurement 
semantics, ensuring that results remain comparable and reproducible across tool 
versions, PDKs, and stack variants.

\begin{figure}[t]
  \centering
  \includegraphics[width=0.85\linewidth]{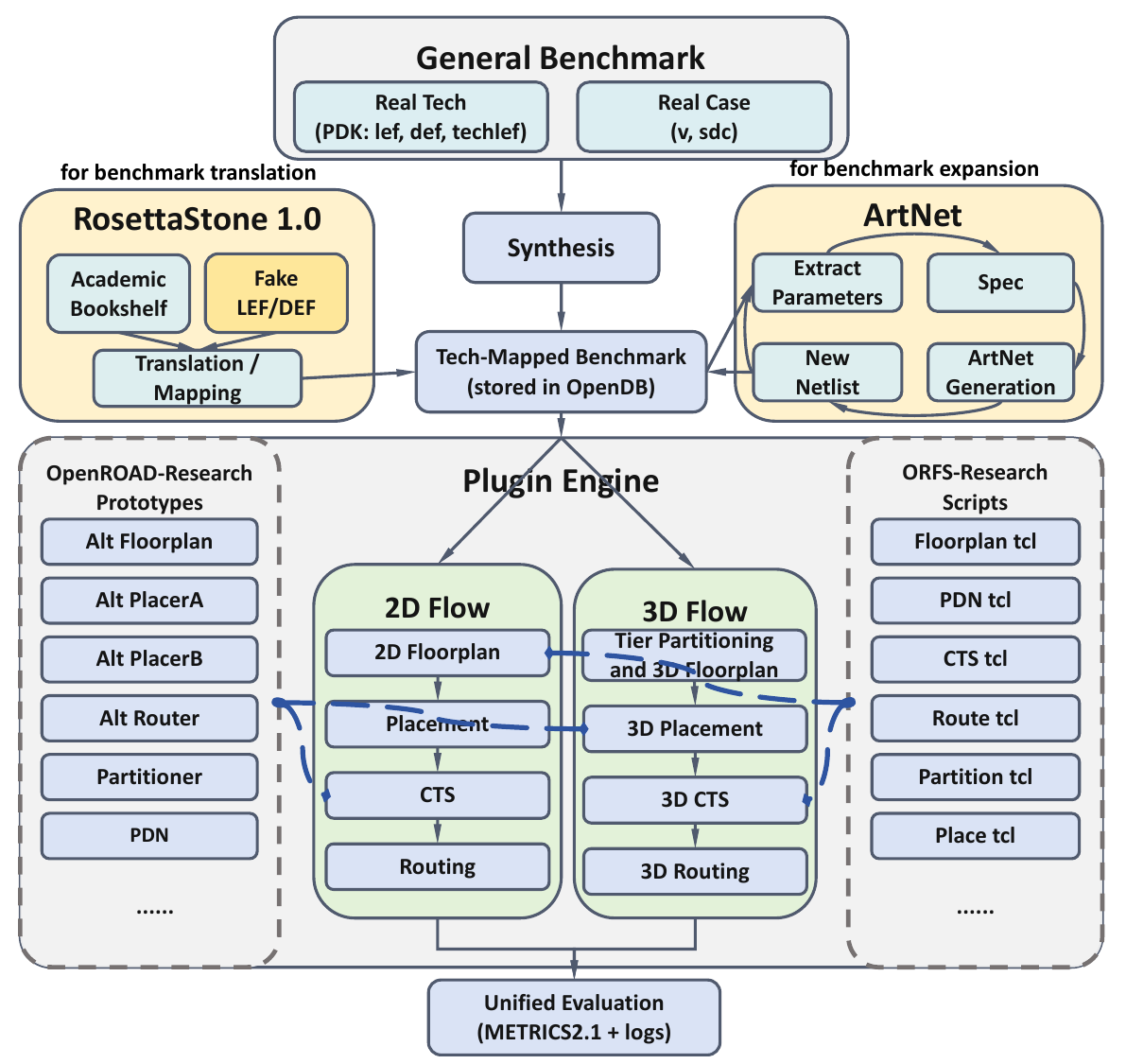}
  \caption{RosettaStone~2.0 roadmap (vision) to sustain and 
  expand the open-source benchmarking backplane. Not all 
  planned components are fully validated in this paper.}
  \label{fig:roadmap}
  \vspace{-1em}
\end{figure}

We frame RosettaStone~2.0 (Figure~\ref{fig:roadmap}) as an advance over 
RosettaStone 1.0 \cite{kahng_rosettastone_2022}. 
Put together, OpenROAD-Research, ORFS-Research, our end-to-end Pin-3D flow
and RosettaStone~2.0 form a concrete instance of a sustainable benchmarking 
backplane. Our goal is for this backplane to enable new algorithms and tool 
prototypes to be integrated, evaluated, and compared under shared conditions, 
supporting credible and cumulative progress in both 2D and 3D physical design research.

\section{Pin-3D Enablement and Flow}
\label{sec:pin3d_flow}

This section presents Pin-3D as an initial, concrete realization of RosettaStone~2.0
within the IEEE CEDA DATC \textit{ORFS-Research} repository.
Our goal is not to claim best-possible QoR, but to demonstrate that 
RosettaStone~2.0 can deliver a maintained, reproducible RTL-to-GDS reference flow 
for F2F hybrid-bonded 3D designs, including versioned 3D enablements, 
stage-wise checkpoints, and METRICS2.1-compatible reporting.
By instantiating Pin-3D on top of OpenROAD-Research and ORFS-Research, we show how 
new 3D methodologies can be packaged, validated, and compared under shared inputs and
measurement semantics.

\subsection{Overview}
\label{subsec:pin3d_overview}

\begin{figure}[t]
  \centering
  \includegraphics[width=0.8\columnwidth]{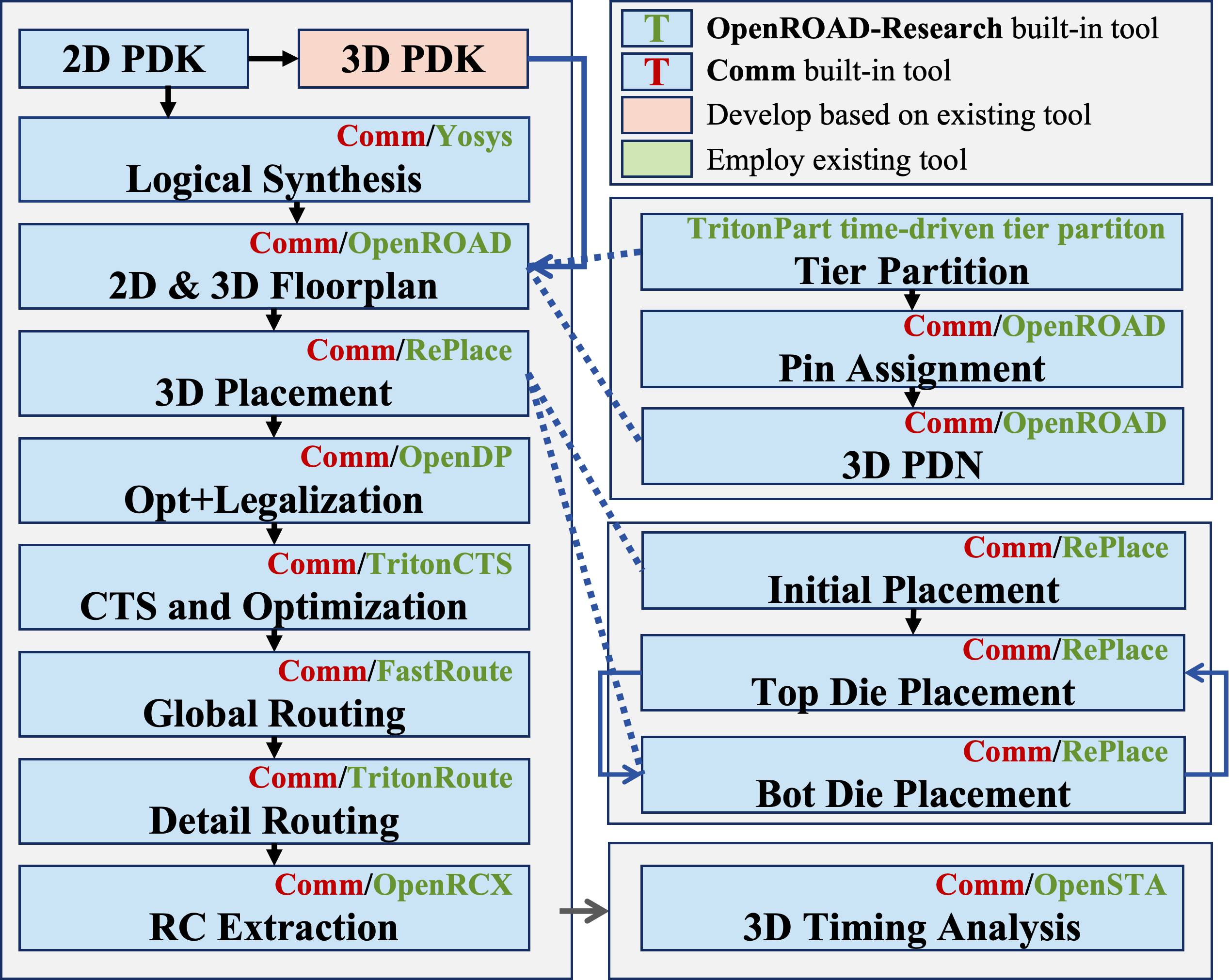}
  \caption{Overview of the Pin-3D flow built on OpenROAD-Research and ORFS-Research.}
  \label{fig:pin3d_flow}
  \vspace{-1em}
\end{figure}

Figure~\ref{fig:pin3d_flow} presents an overview of our homogeneous and
heterogeneous Pin-3D flow, implemented on top of OpenROAD-Research and
ORFS-Research. The flow comprises the following key stages:
\begin{itemize}[noitemsep, topsep=0pt, leftmargin=*]
\item \textbf{Synthesis and 2D abstraction:} RTL is synthesized into a flat
gate-level netlist using a common logical library, which is later mapped to
tier-specific physical libraries for heterogeneous designs.

\item \textbf{Partitioning and 3D floorplanning.} An initial 2D floorplan
undergoes timing-driven bipartitioning~\cite{zhiang_tritonpart_2023} to assign
cells to tiers. The design is then translated into tier-aware 3D views with
independent power delivery networks.

\item \textbf{Iterative alternating-tier 3D placement.}
Placement refinement begins from a global 2D placement and proceeds by alternating
optimization between tiers: one tier is fixed (loaded as \textbf{COVER}) while the other 
is optimized for timing/congestion, followed by tier-aware legalization; the roles are 
then swapped and iterated.

\item \textbf{3D clock tree synthesis (CTS).} The clock tree is built on the
bottom tier. Top-tier sinks connect to this tree through inter-tier vias,
leveraging the unified 2D representation for vertical connectivity.

\item \textbf{3D routing and optimization.} Routing utilizes the full metal
stack, modeling hybrid bonding terminals as special vias. Parasitics are then
extracted to drive post-route timing and power closure.

\item \textbf{Metrics collection:} The flow records runtime, memory, timing,
power, wirelength, congestion, and violations (DRVs/FEPs) in a structured
format for both open-source and commercial reference flows.
\end{itemize}

\noindent
We highlight three key features of our Pin-3D-style framework:
\begin{itemize}[noitemsep, topsep=0pt, leftmargin=*]
\item \textbf{Unified 2D technology abstraction for F2F 3D.}
We model HBTs as special vias in an extended 2D metal stack, enabling unmodified 2D routers to realize cross-tier connections.

\item \textbf{Iterative tier-by-tier optimization and legalization with 2D placers.}
We alternate placement optimization between tiers using \textbf{COVER} views, with tier-aware legalization to enforce tier-specific sites and masters for homogeneous and heterogeneous stacks.

\item \textbf{End-to-end 3D reference flow with hybrid toolchain support.}
A unified scripting interface runs the RTL-to-GDS flow using OpenROAD, commercial tools, or mixed toolchains.
Stage-wise checkpoints and METRICS2.1-compatible reports are exported to
enable consistent, reproducible measurement.
\end{itemize}


Throughout the flow, constraints and 3D enablements are aligned under an
explicit evaluation contract.
All metrics are reported using a consistent schema to eliminate ambiguity in
metric definitions and measurement points.

\subsection{PDK Preparation}
\label{subsec:pdk_prep}


Process Design Kit (PDK) preparation is a foundational component of a 
robust 3D physical design flow. A 3D PDK must support stable and automated 
execution, while still capturing the essential integration effects
needed for meaningful tool and algorithm evaluation.

\begin{figure}[t]
  \centering
  \includegraphics[width=0.95\linewidth]{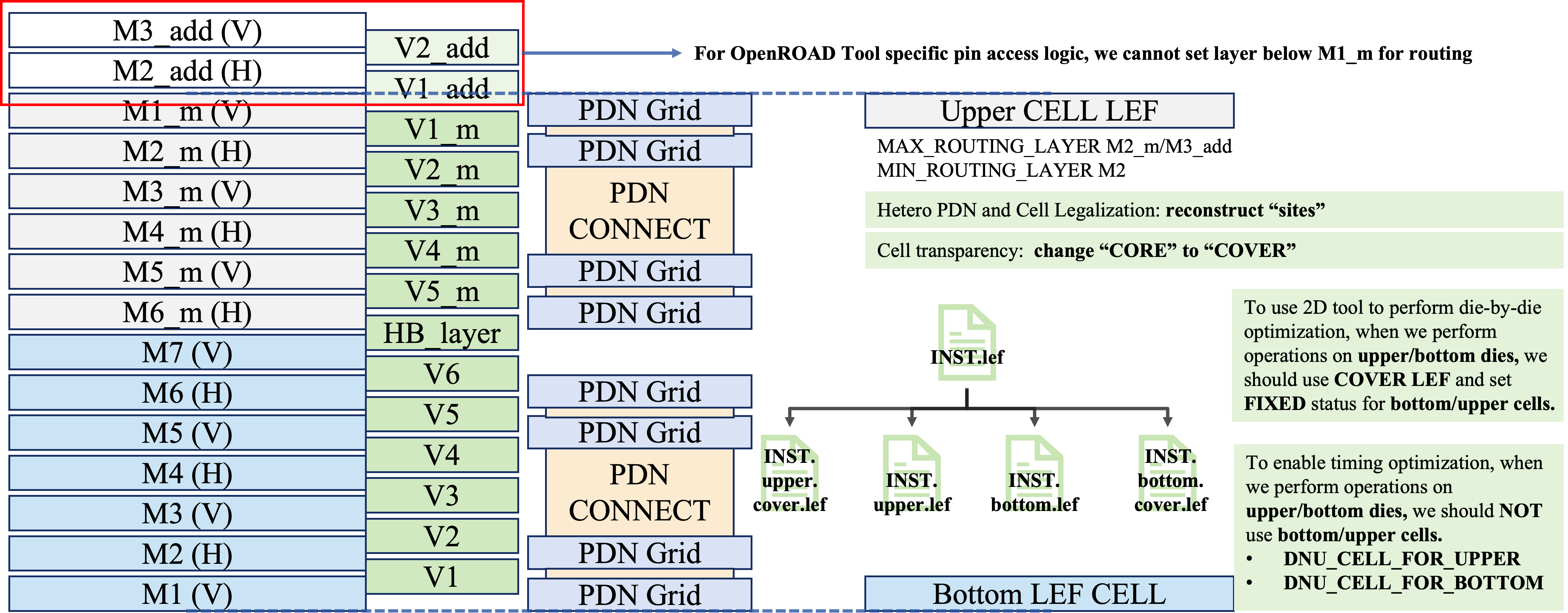}
  \caption{Metal stack, PDN strategy, and tier strategy in the 3D ASAP7 PDK.}
  \label{fig:ASAP7_3D_PDK}
  \vspace{-1em}
\end{figure}

\begin{figure}[t]
  \centering
  \begin{subfigure}{0.43\columnwidth}
    \includegraphics[height=3.2cm, keepaspectratio]{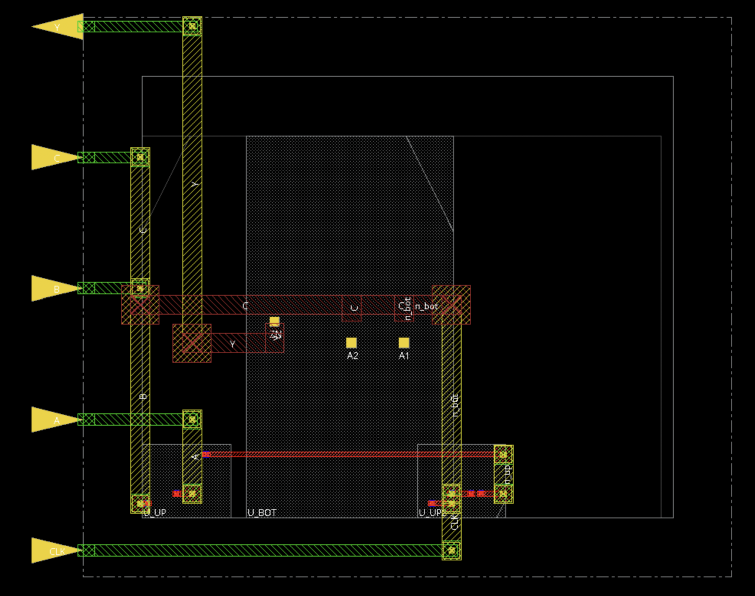}
  \end{subfigure}
  \hspace{0.05\columnwidth}
  \begin{subfigure}{0.43\columnwidth}
    \includegraphics[height=3.2cm, keepaspectratio]{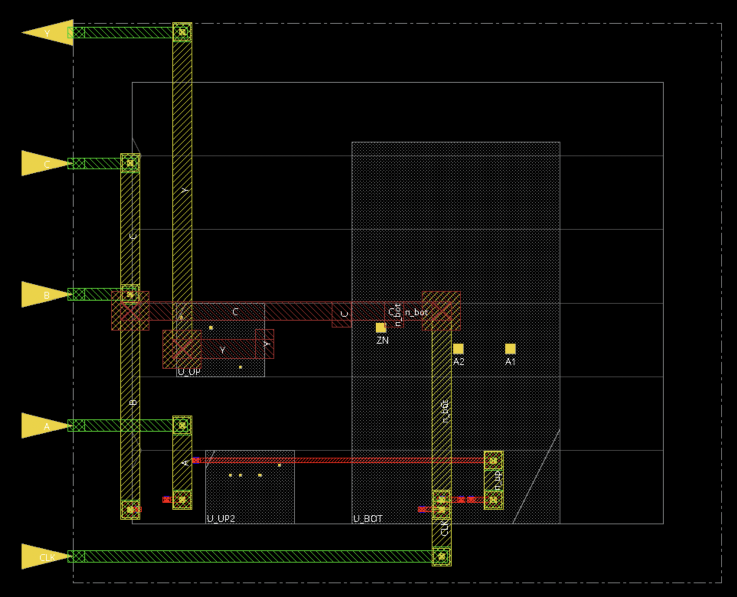}
  \end{subfigure}
  \caption{Rebuilding of rows for heterogeneous legalization.}
  \label{fig:RebuildRow}
  \vspace{-1em}
\end{figure}

Figure~\ref{fig:ASAP7_3D_PDK} summarizes our 3D ASAP7 enablement, which is
derived from the standard ASAP7 PDK~\cite{vashishtha_asap7_2017, asap7_repo}. 
We construct a 
face-to-face stack by replicating the 2D metal stack per tier and introducing 
a dedicated cut layer to model hybrid bonding terminals (HBTs) as vertical vias. 
\texttt{M2\_add} and \texttt{M3\_add} are compatibility layers required by the legacy 
ORFS-Research router's pin-access limitations (i.e., upper-only pin access); this will 
be removed once the router is updated to support lower-metal pin access, consistent 
with the COMM TechLEF.
Since the two dies in an F2F configuration are fabricated independently prior to bonding~\cite{liu_routing-aware_2024},
we implement isolated power delivery networks (PDNs) on the two tiers. 
This design enables independent supplies and voltage
domains, which is important for supporting heterogeneous stacks.

To support 3D physical implementation, we construct 3D standard-cell libraries by deriving tier-specific physical
views from a base 2D library. For each logical cell, we generate distinct LEF
masters for the bottom and top tiers (denoted by \textbf{\_bottom} and \textbf{\_upper}
suffixes) by reassigning the cell's metal layers to the appropriate tier. 
We additionally provide \textbf{COVER} LEF views for both tiers, which serve
as physical abstractions to exclude the inactive tier from overlap and density
calculations while preserving connectivity~\cite{pentapati_pin-3d_2024}.

For homogeneous stacks, logic synthesis targets the standard 2D library;
the resulting netlist is later remapped to tier-specific masters
following the partitioning stage.
For heterogeneous stacks, we synthesize against a unified library that represents 
the intersection of available cells across the two technologies.
To ensure compatibility, any physical pins not shared between the tiers are
hidden as internal pins in this logical view, allowing the synthesized
netlist to be flexibly mapped to either tier.
Although this intersection method keeps the flow simple, hiding of pins that 
are not shared requires careful checking that circuit function is preserved.
It might also limit optimization scope, compared to natively utilizing multiple libraries.
During the physical design stage, logical cells are assigned to specific tiers,
and legality is ensured by reconstructing the rows with tier-specific sites (Figure~\ref{fig:RebuildRow}).

\subsection{Physical Design Flow Details}
\label{subsec:flow_details}

This subsection details the script-based implementation of each stage in
Figure~\ref{fig:pin3d_flow}. Two design choices enable a standard 2D toolchain
to operate on 3D designs: (i) a unified technology representation that models
inter-tier connections as vias on an HBT cut layer, and (ii) a \textbf{tier
strategy} that alternates optimization between tiers by loading \textbf{COVER}
views for the inactive tier and restricting cell usage to the active tier.

\medskip
\noindent\textbf{Stage 1: Synthesis and technology-neutral netlist.}
We synthesize (using Yosys or a commercial tool) a flat gate-level netlist that 
is input to
timing-driven bipartitioning. Homogeneous designs use the native 2D library, 
while heterogeneous designs use the unified logical library described in
Section~\ref{subsec:pdk_prep}, producing a single netlist that is later mapped
to tier-specific physical masters.

\medskip
\noindent\textbf{Stage 2: Floorplan, partitioning, and tier view generation.}
We first generate an initial 2D floorplan based on target utilization and aspect ratio.
The floorplan and netlist are then passed to TritonPart for timing-driven bipartitioning.
For heterogeneous stacks (ASAP7+NanGate45\_3D in this paper), we additionally enable 
a capacity-aware baseline split to reflect technology-dependent standard-cell area differences.
More precisely: we invoke TritonPart with a partition balance constraint, 
denoted as \texttt{UBfactor} (unbalance factor).
After sweeping this constraint from \texttt{PAR\_BAL\_LO} to \texttt{PAR\_BAL\_HI}, 
we adopt the minimum cutsize solution found. This cutsize serves as an early 
estimate for HBTs.\footnote{We implement two independent sweep modes, each evaluated 
over multiple \texttt{PAR\_BAL\_ITERATION} points (11 by default), and select the 
minimum-cut solution in timing-driven mode.
\emph{(A) UBfactor sweep.}
We sweep the imbalance constraint (\texttt{UBfactor}) uniformly from 
\texttt{PAR\_BAL\_LO} to \texttt{PAR\_BAL\_HI}, run timing-driven bipartitioning at 
each point, and select the solution with the minimum cutsize.
\emph{(B) Base-balance sweep.}
To account for unequal effective placement capacities across tiers, we sweep a 
technology-aware \texttt{base\_balance} while keeping timing-driven partitioning enabled.
For ASAP7+NanGate45, we start from \texttt{base\_balance}=$(0.06,0.94)$
(reflecting $\sim$16$\times$ standard-cell area ratio) and linearly reduce the 
imbalance over 11 points toward a less skewed split; at each point, we run
TritonPart and pick the minimum-cut solution.
In both modes, the selected cutsize provides an early estimate of cross-tier 
connectivity (hybrid bonding terminals).}

\begin{figure}[t]
  \centering
  \includegraphics[width=0.9\columnwidth]{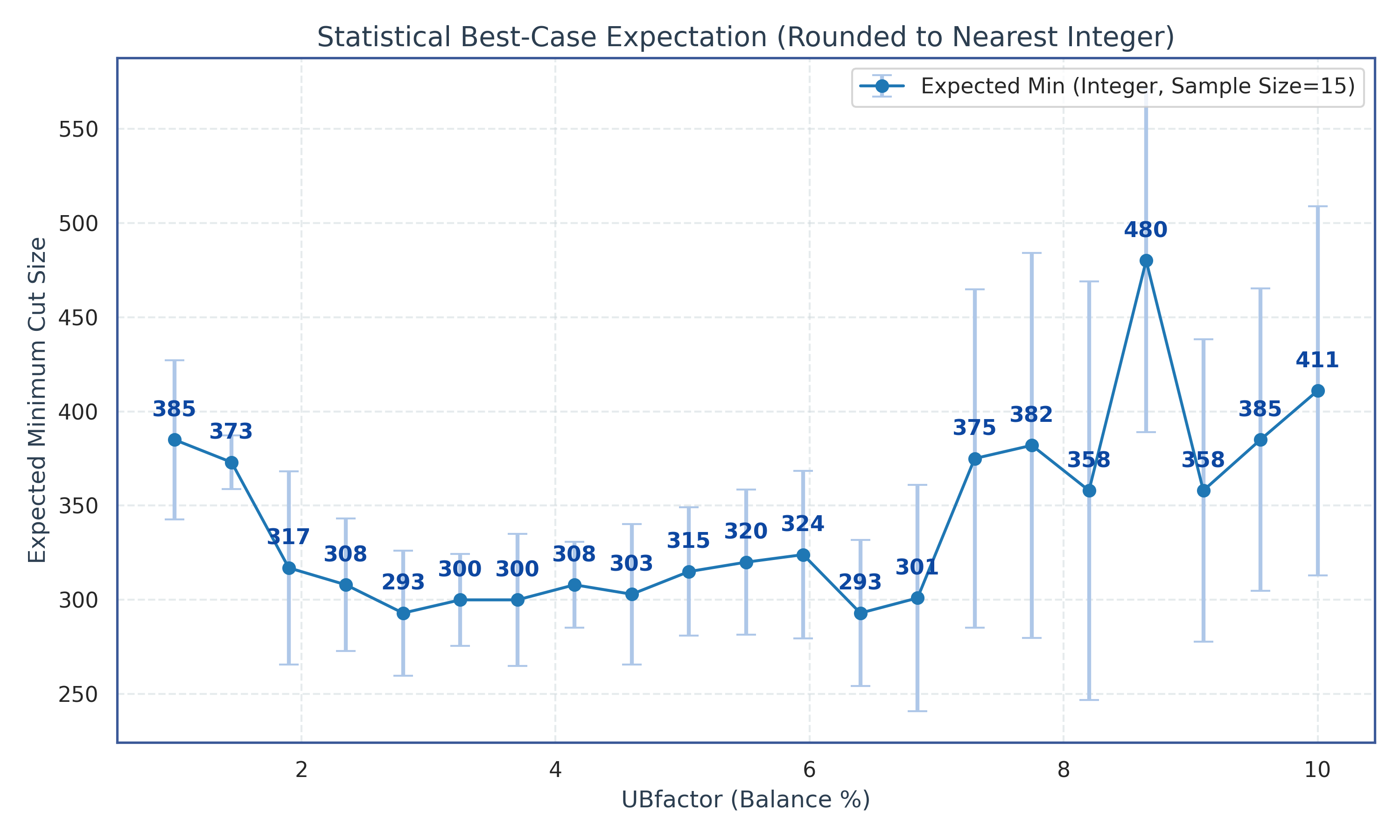}
  \caption{Cross-tier net count vs. \texttt{UBfactor} based on
  50 random seeds per point. The trace shows the \textbf{Expected Minimum Cutsize}
  (bootstrapped, $N=15$), representing the typical best result achievable.}
  \label{fig:ub_vs_cutsize}
  \vspace{-1em}
\end{figure}

We then generate tier-specific DEF/Verilog views through a conversion step:
(i) remapping logical masters to tier-specific physical masters
(\textbf{\_bottom}/\textbf{\_upper}),
(ii) applying pin remapping for heterogeneous designs, and
(iii) tying off pins that are not present in the logical abstraction.

For homogeneous stacks (7+7 and 45+45), we assign IO pins on the bottom
tier. For the heterogeneous stack (7+45), we assign IO pins on the top tier,
since it hosts more cells and this choice empirically reduces cross-tier
connections.

\begin{figure*}[t]
  \centering
  \captionsetup[subfigure]{font=scriptsize, labelfont=bf, justification=centering}

  \begin{subfigure}[b]{0.47\columnwidth}
    \centering
    \includegraphics[width=\linewidth]{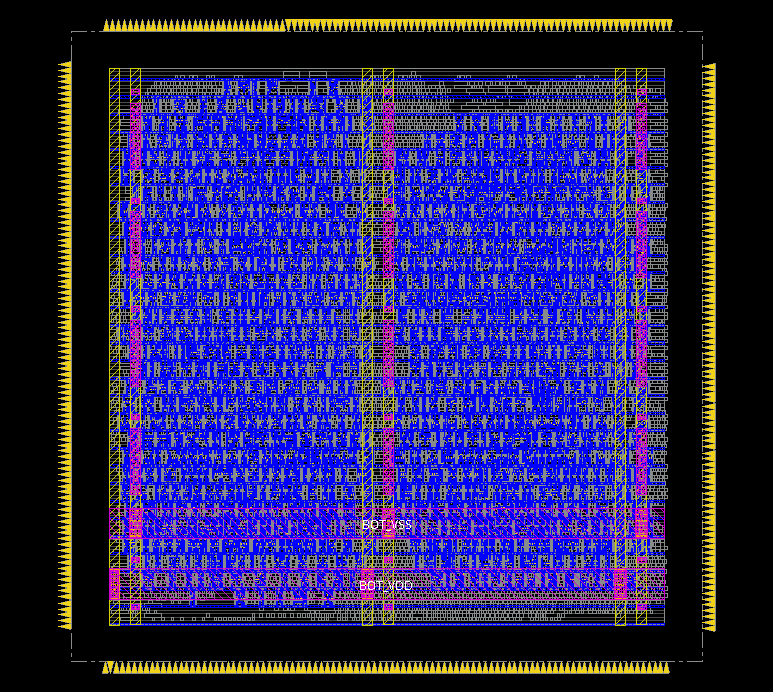}
    \caption{Bottom PDN (M1 stripe in blue)}
    \label{fig:bot_pdn}
  \end{subfigure}\hfill
  \begin{subfigure}[b]{0.47\columnwidth}
    \centering
    \includegraphics[width=\linewidth]{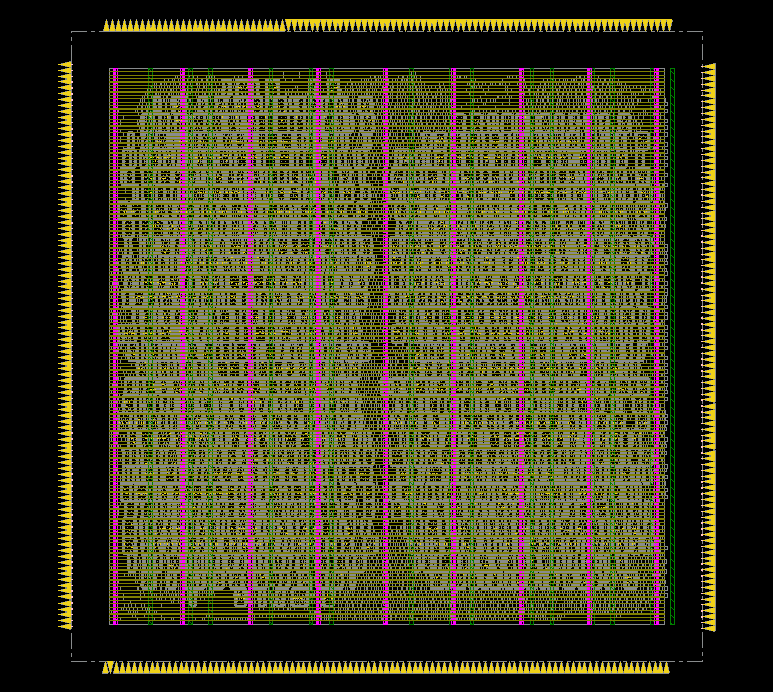}
    \caption{Top PDN (M1\_m stripe in yellow)}
    \label{fig:top_pdn}
  \end{subfigure}\hfill
  \begin{subfigure}[b]{0.47\columnwidth}
    \centering
    \includegraphics[width=\linewidth]{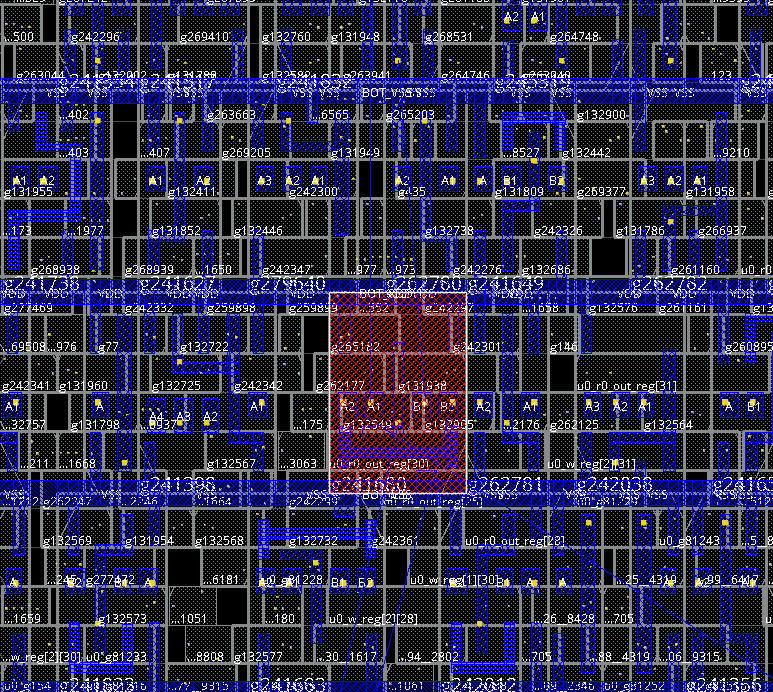}
    \caption{Bottom cells (in blue)}
    \label{fig:bot_cell}
  \end{subfigure}\hfill
  \begin{subfigure}[b]{0.47\columnwidth}
    \centering
    \includegraphics[width=\linewidth]{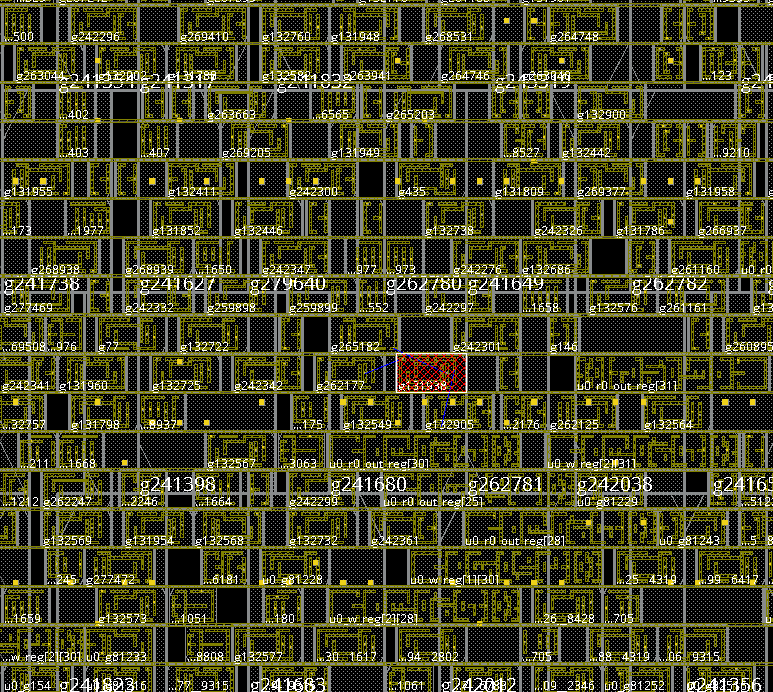}
    \caption{Top cells (in yellow)}
    \label{fig:top_cell}
  \end{subfigure}
  
  \vspace{-1em}
  \caption{Heterogeneous PDN grid and cell placement (\textit{aes}, 7+45).}
  \label{fig:bot_top_PDN_Cell}
\end{figure*}

Finally, we build electrically isolated PDNs for each tier using symmetric grid
topologies (Figure~\ref{fig:bot_top_PDN_Cell}).

\medskip
\noindent\textbf{Stage 3: Iterative 3D placement with adaptive strategy.}
In addition to loading \textbf{COVER} views, we evaluate two tier strategies for
placement and legalization in both homogeneous and heterogeneous stacks:
(i) a \textbf{Restricted} strategy, which limits available masters to the active
tier and keeps the inactive tier fixed; and
(ii) a \textbf{Flexible} strategy, which does not enforce tier constraints and
allows the placer to choose between tier-specific masters during optimization.

Our reference flow uses the \textbf{Flexible} strategy, while we also study the
\textbf{Restricted} strategy and report WNS and total power.
The subsequent resizing and legalization stage is always \textbf{Restricted}, to prevent
buffers from being inserted onto the inactive tier and creating tier-illegal
solutions.

\begin{figure*}[t]
  \centering
  \captionsetup[subfigure]{font=scriptsize, labelfont=bf, justification=centering}
  
  \begin{subfigure}[b]{0.47\columnwidth}
      \centering
      \includegraphics[width=\linewidth]{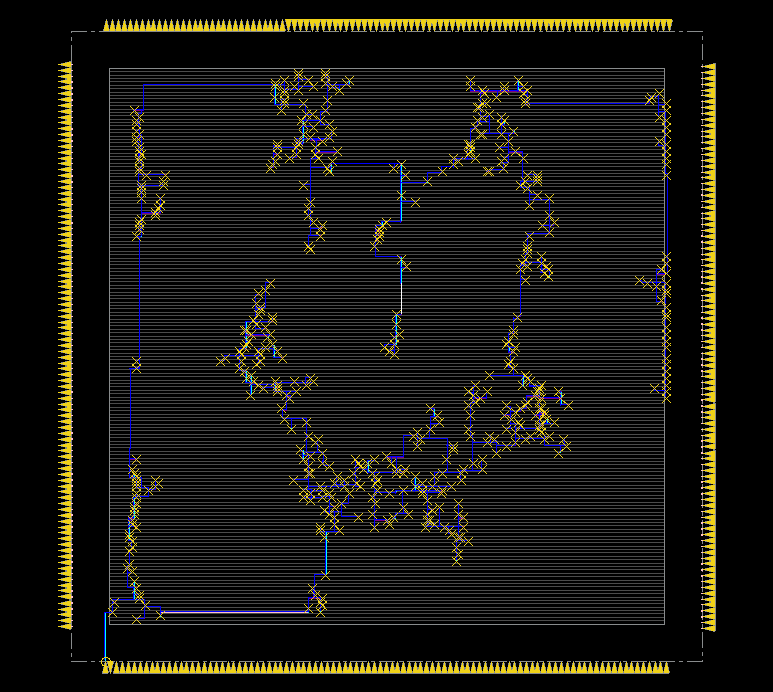}
      \caption{Top die clock tree \& routed clock net}
      \label{fig:CLK_Net}
  \end{subfigure}\hfill
  \begin{subfigure}[b]{0.47\columnwidth}
      \centering
      \includegraphics[width=\linewidth]{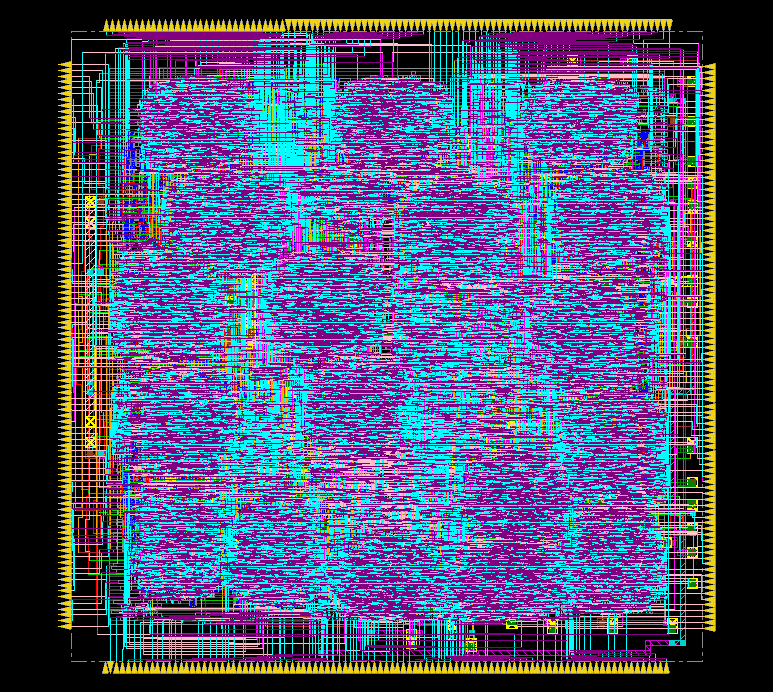}
      \caption{Post-route optimized layout}
      \label{fig:route_signal}
  \end{subfigure}\hfill
  \begin{subfigure}[b]{0.47\columnwidth}
      \centering
      \includegraphics[width=\linewidth]{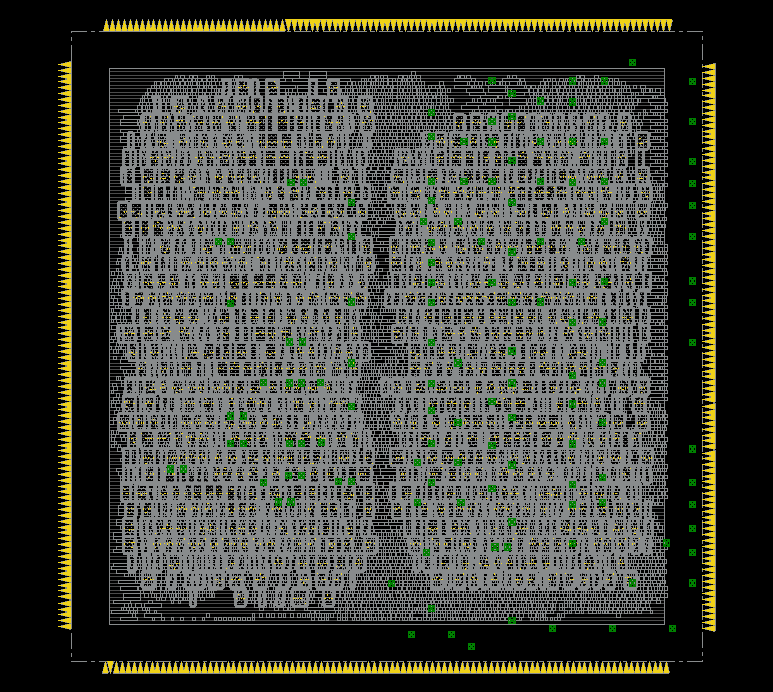}
      \caption{HBTs assigned by the router (green)}
      \label{fig:hbt}
  \end{subfigure}\hfill
  \begin{subfigure}[b]{0.47\columnwidth}
      \centering
      \includegraphics[width=\linewidth]{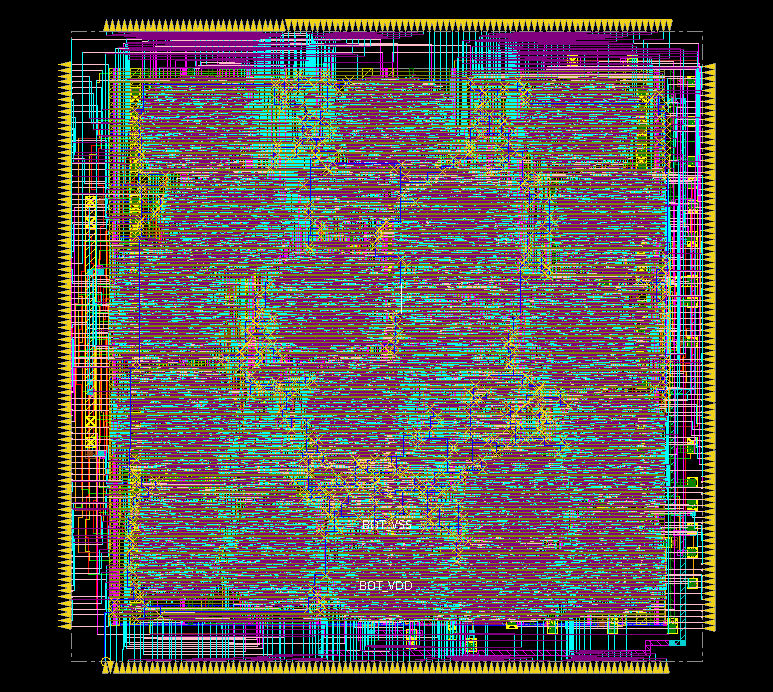}
      \caption{Heterogeneous 3D \textit{aes}}
      \label{fig:final_layout}
  \end{subfigure}
  
  \vspace{-1em}
  \caption{HBTs and routed signal nets (\textit{aes}, 7+45).}
  \label{fig:postPR}
\end{figure*}

\medskip
\noindent\textbf{Stage 4: 3D clock tree synthesis.}
We perform CTS on a specific active tier (\textbf{bottom} for homogeneous stacks,
\textbf{top} for the 7+45 stack) while holding the other tier fixed and utilize 
the {\bf Restricted} strategy
according to the selected tier. Clock sinks on the other tier are connected
through inter-tier vias, leveraging the unified cut-layer abstraction. For the
ORD flow, we run an additional legalization step after buffer insertion.
Figure~\ref{fig:CLK_Net} shows the resulting clock tree and routed clock net.

\medskip
\noindent\textbf{Stage 5: 3D routing and parasitic extraction.}
Global and detailed routing are executed with standard 2D routers. Cross-tier
connections are realized by inserting HBT vias defined in the unified tech LEF.
After routing, we run parasitic extraction using the same unified tech LEF,
producing SPEF that includes inter-tier RC for final timing and verification
(Figure~\ref{fig:postPR}).

\medskip
\noindent\textbf{Stage 6: Metrics collection and reporting.}
We collect QoR metrics at multiple checkpoints. For the OpenROAD tool,
ORFS-Research reporting commands emit runtime, memory, wirelength, timing, 
and violations in METRICS2.1-style structured JSON. For COMM, we run an equivalent set of
built-in reports and parse them into CSV with consistent metric definitions.
All scripts, inputs, and outputs are version-controlled to ensure reproducible
evaluation.


\section{Pin-3D Validations}
\label{sec:pin3d_validation}

This section reports experimental studies that showcase how
the Pin-3D flow and ORFS-Research can enable fair and reproducible
physical-design (PD) comparisons across tools and algorithms. We report a
range of results from both the open-source flow (ORD) and a commercial 
reference flow (COMM) to demonstrate portability, flexibility, and end-to-end 
execution; absolute quality-of-results (QoR) measurement is not a primary 
objective of this work. All experiments are run on a Linux server with an 
Intel Xeon Gold 6230N CPU and 503\,GB RAM.

\begin{figure}[h]
    \centering
    \includegraphics[width=\linewidth]{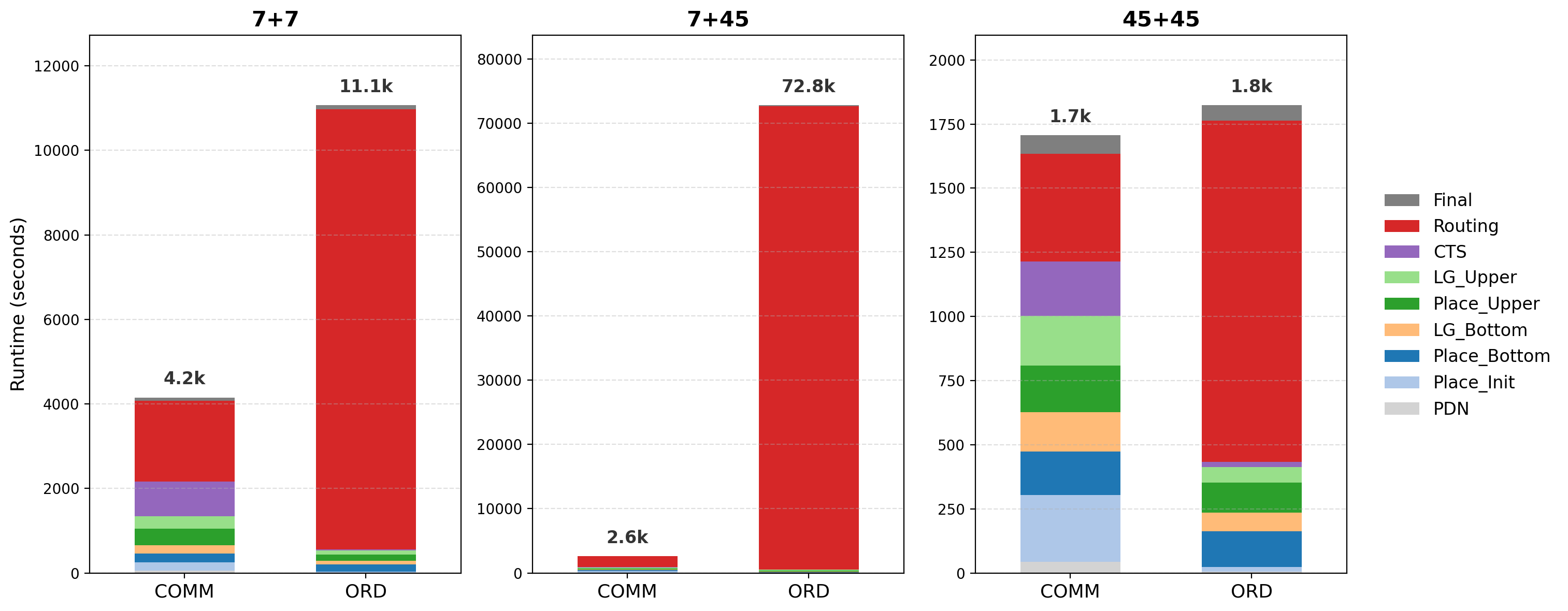}
    \caption{Runtime breakdowns for jpeg. Numbers above the bars indicate total runtime 
    in seconds.}
    \label{fig:jpeg_runtime_analysis}
\end{figure}

\begin{figure}[!htbp]
  \centering
  \begin{subfigure}[b]{0.48\linewidth}
      \centering
      \includegraphics[width=\linewidth]{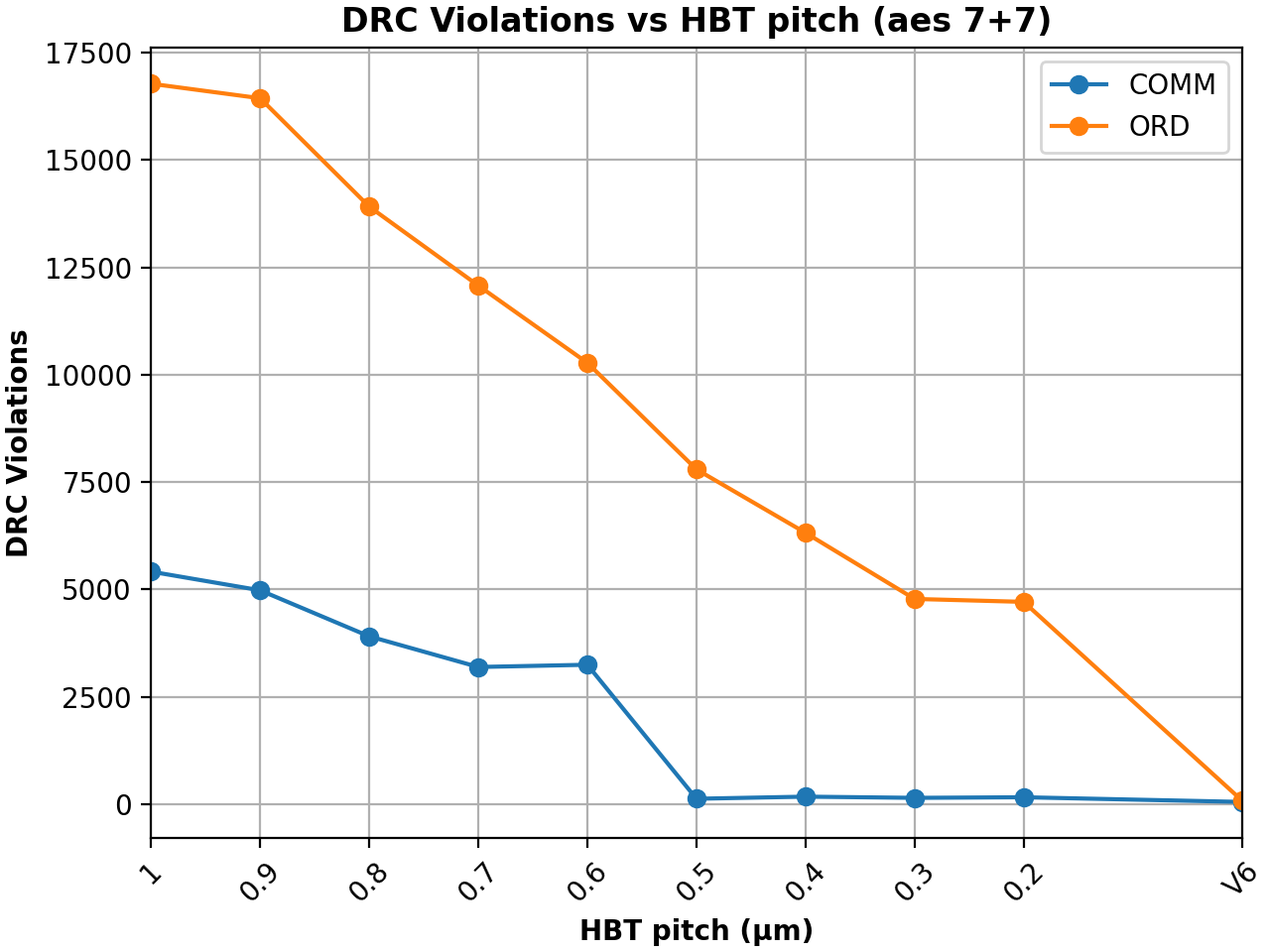}
  \end{subfigure}
  \hfill
  \begin{subfigure}[b]{0.48\linewidth}
      \centering
      \includegraphics[width=\linewidth]{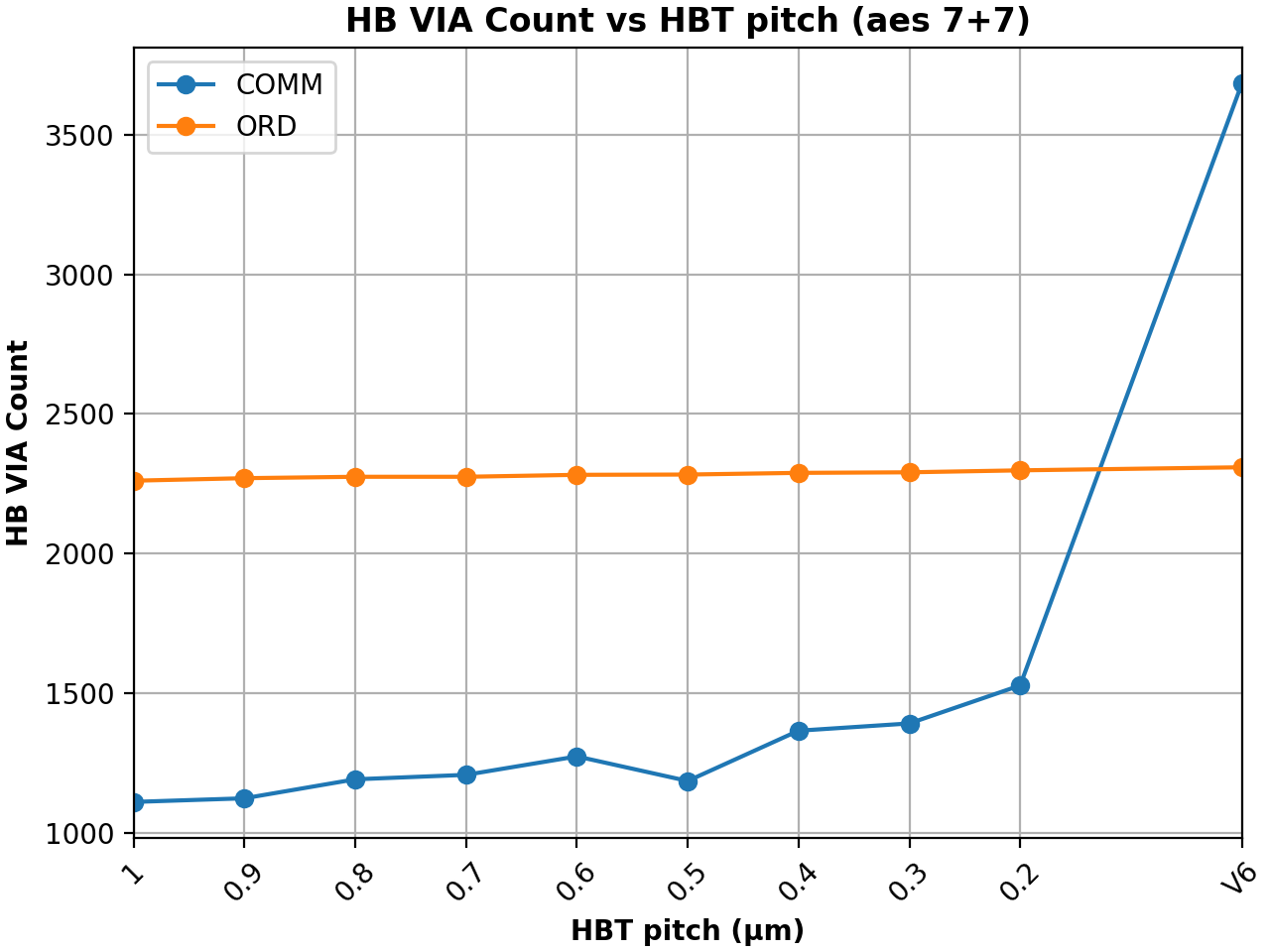}
  \end{subfigure}
  \caption{Impact of HBT pitch scaling on routing closure: DRV count (left) and HBT via count (right) versus HBT pitch.}
  \label{fig:hb_pitch_analysis}
\end{figure}

\begin{figure}[h]
  \centering
  \includegraphics[width=\linewidth]{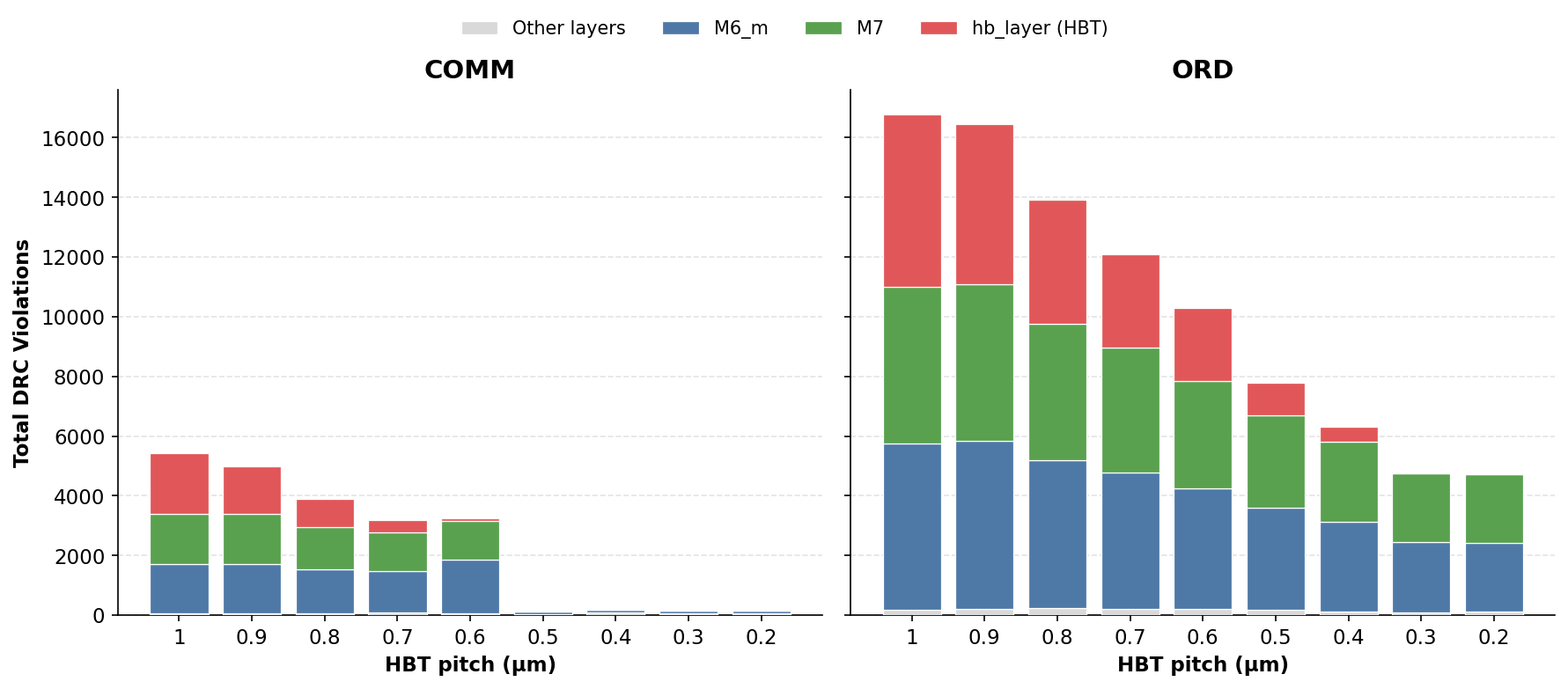}
  \caption{Impact of HBT pitch on DRVs. 
  As the HBT pitch decreases, HBT-related errors (colored) vanish.}
  \label{fig:hb_pitch_drc}
\end{figure}

\begin{figure}[!htbp]
  \centering
  \begin{tabular}{cc}
    \includegraphics[width=0.47\linewidth]{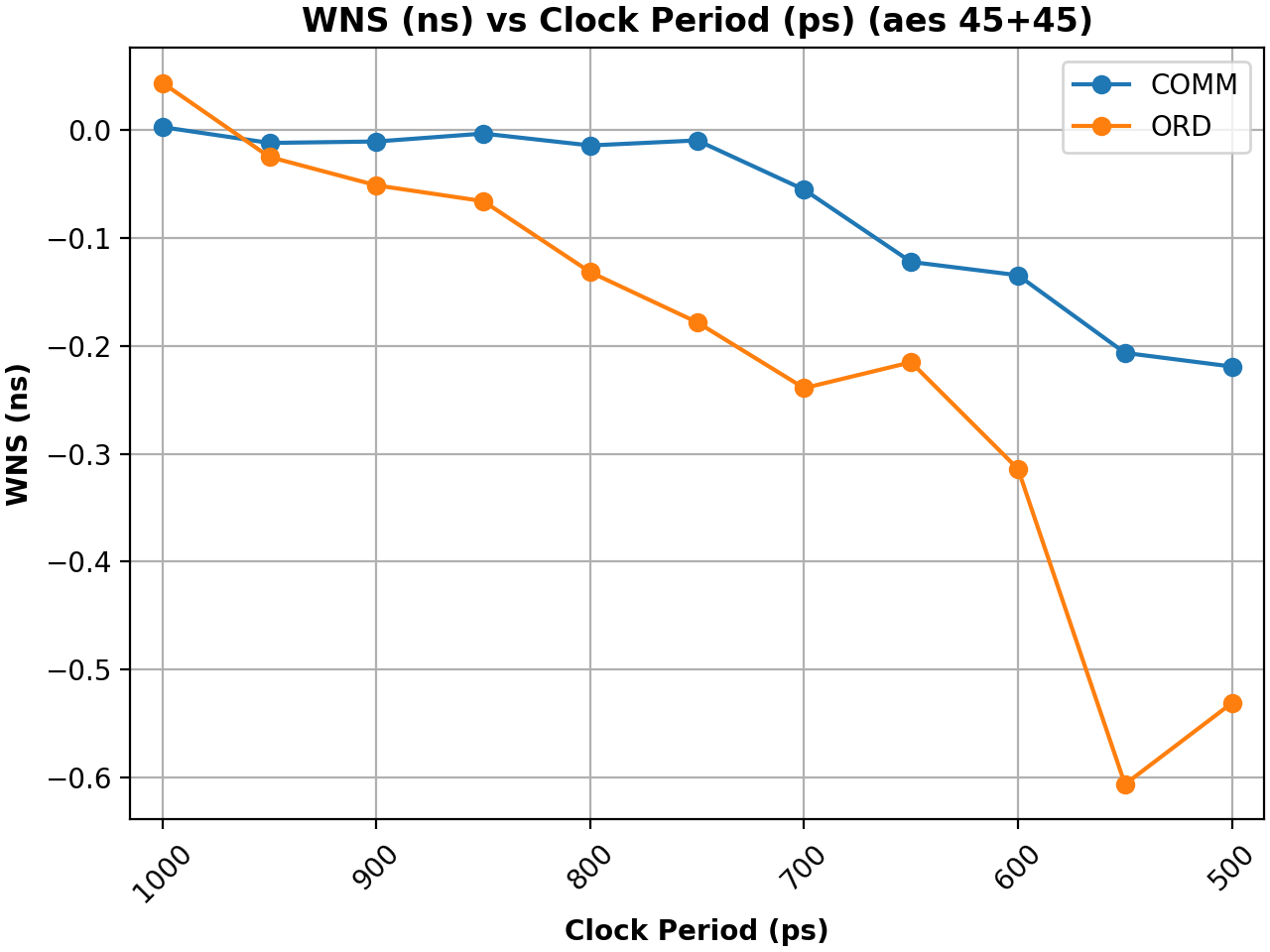} &
    \includegraphics[width=0.47\linewidth]{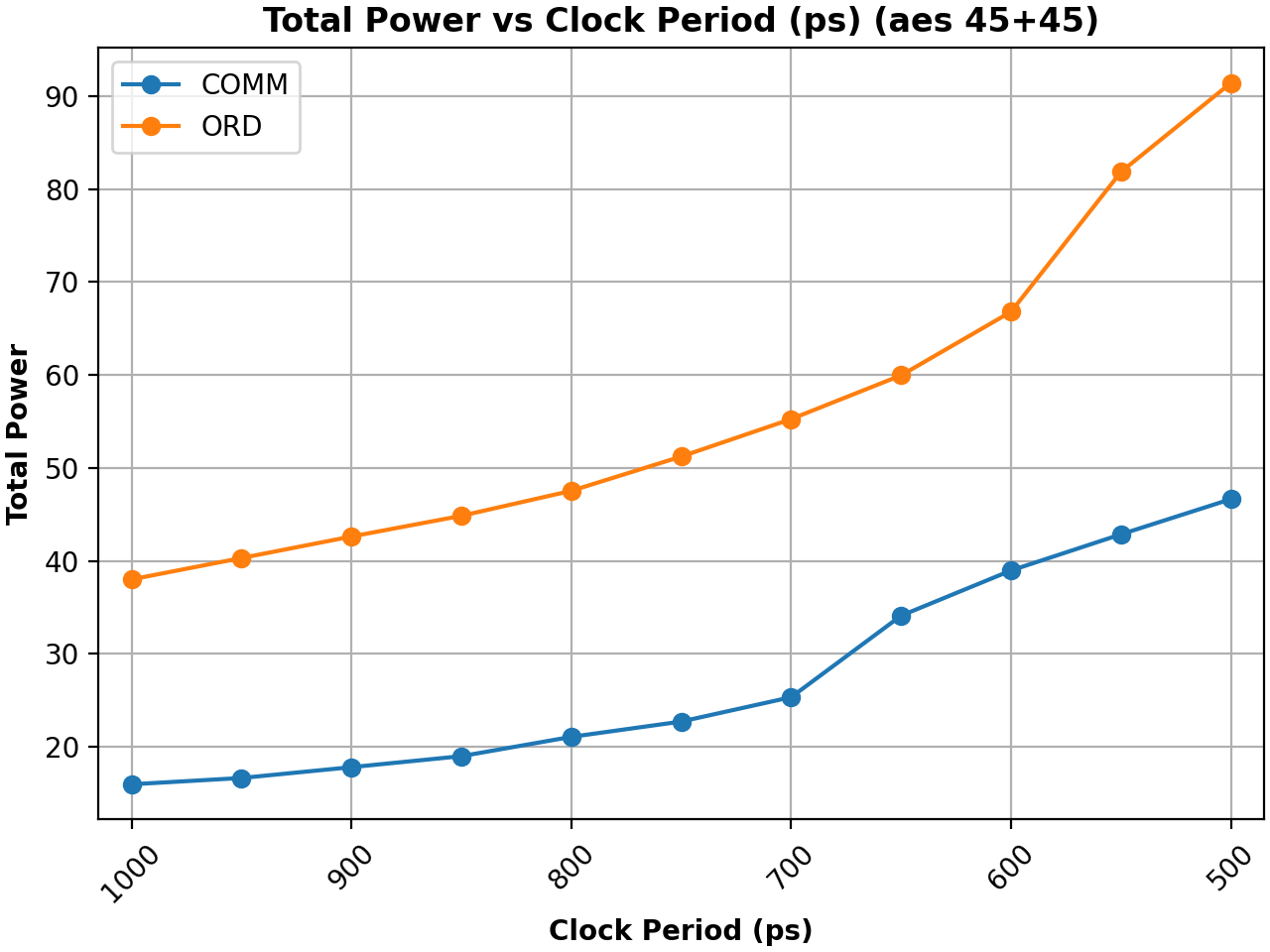} \\
    \includegraphics[width=0.47\linewidth]{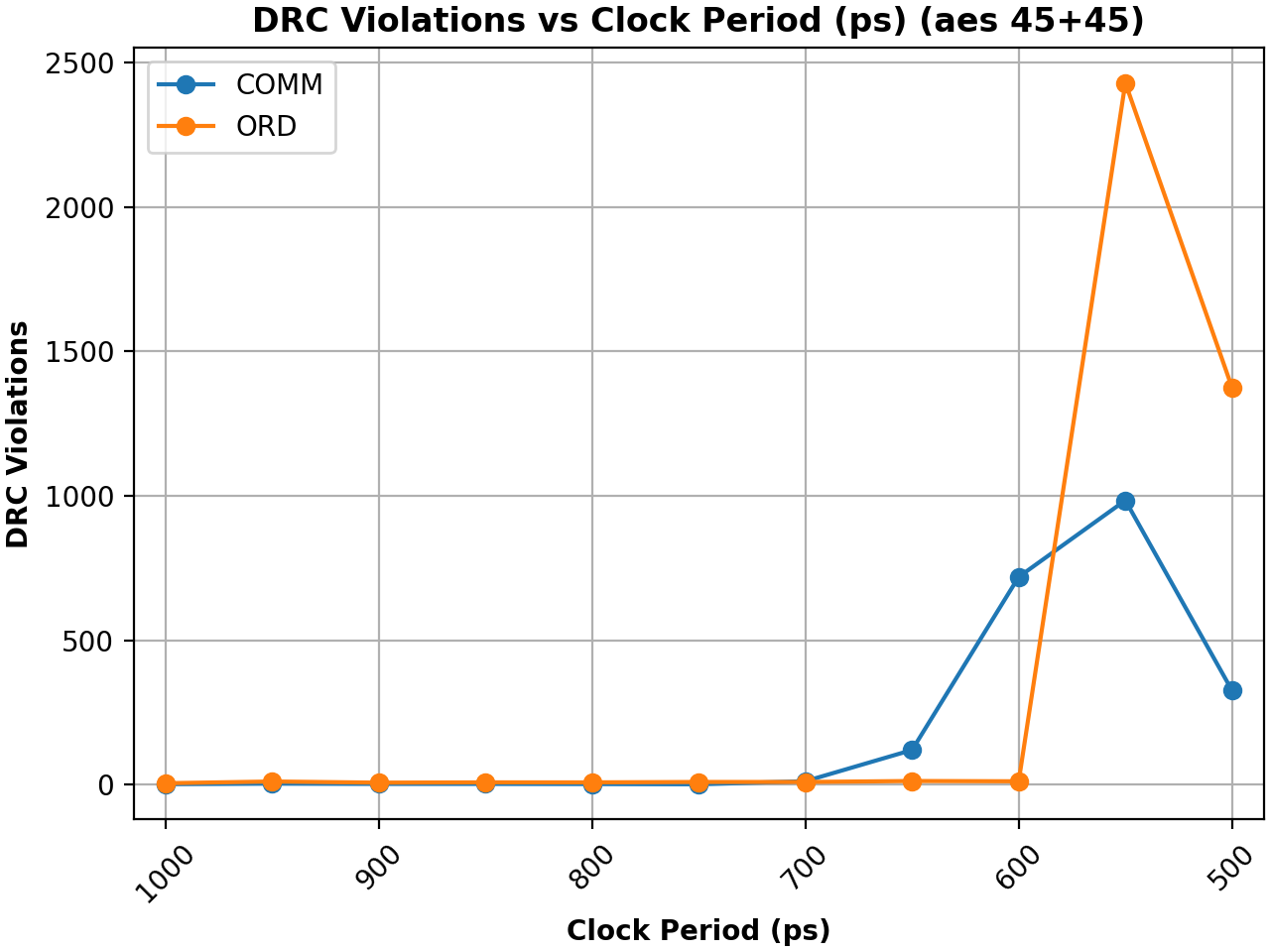} &
    \includegraphics[width=0.47\linewidth]{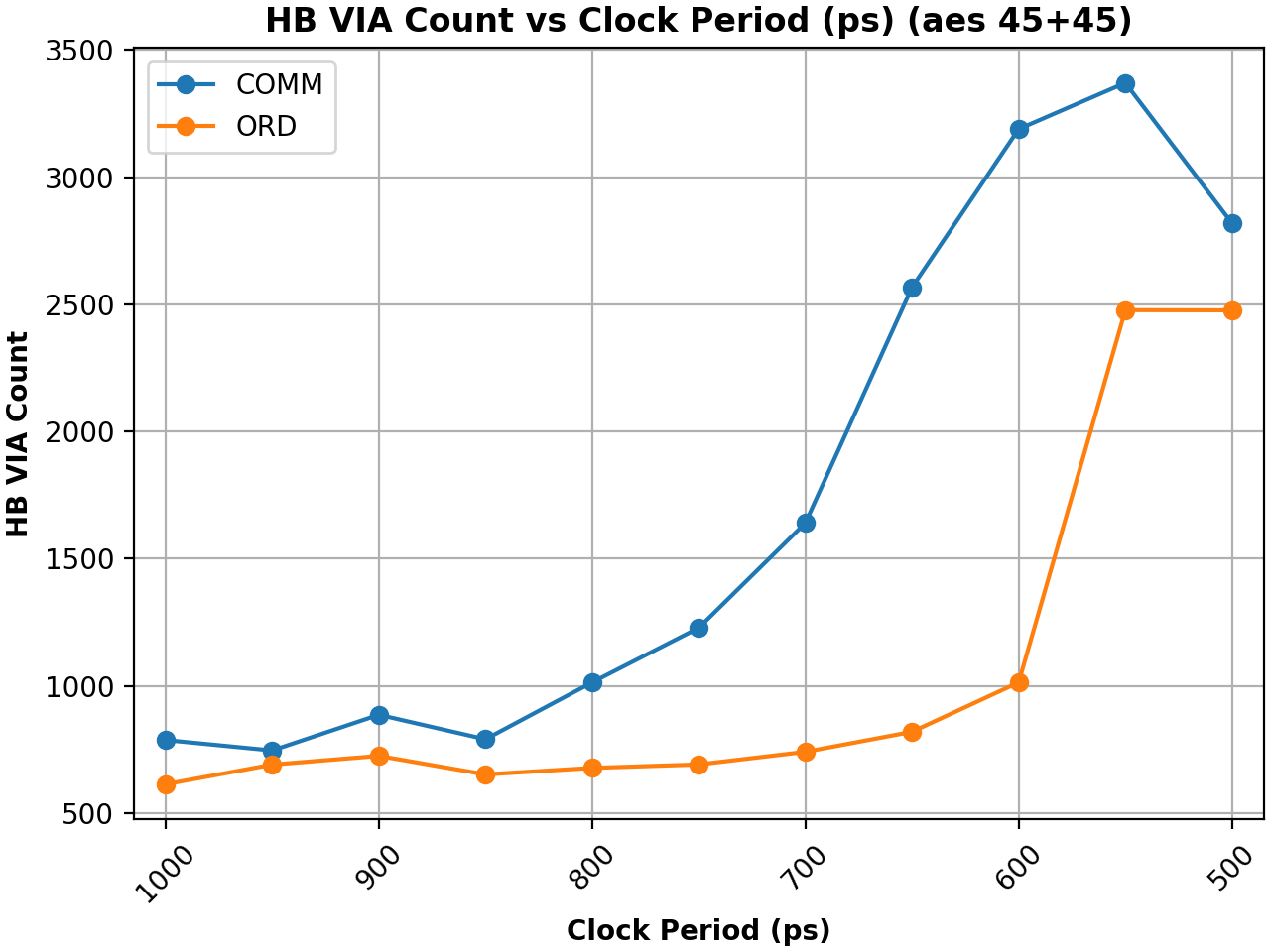}
  \end{tabular}
  \caption{Impact of clock constraint on QoR (\textit{aes}, 45+45).}
  \label{fig:clock_period_analysis}
\end{figure}

We conduct the following experiments:
\begin{itemize}[noitemsep, topsep=0pt, leftmargin=*]
  \item \textbf{3D/2D baseline studies:} three designs across multiple 
  enablements (Tables~\ref{tab:pin_3d_results} and~\ref{tab:2d_results}).
  \item \textbf{Tier strategy comparison:} \textbf{Restricted} vs.\ \textbf{Flexible}
  (Table~\ref{tab:tier_strategy_comparison}).
  \item \textbf{Mixed-toolchain ablations:} swapping synthesis and backend components
  to localize QoR gaps (Table~\ref{tab:synthesis_impact}).
  \item \textbf{Runtime studies:} stage-wise breakdowns (Figure~\ref{fig:jpeg_runtime_analysis}).
  \item \textbf{Sensitivity studies:} (i) HBT pitch
  (Figures~\ref{fig:hb_pitch_analysis} and~\ref{fig:hb_pitch_drc}); and
  (ii) clock period (Figure~\ref{fig:clock_period_analysis}).
\end{itemize}

We use OpenROAD-Research (v2025.11) as the open-source implementation flow (\textbf{ORD}),
and a commercial RTL-to-GDS implementation flow as an external reference baseline (\textbf{COMM}).
To minimize evaluator-induced discrepancies in reported outcomes,
we use a leading commercial P\&R tool, \textbf{Cadence Innovus} (v21.39), as a
common \emph{evaluator} to report timing, power, and physical violations (DRC/DRV/FEP) 
for \emph{all} experiments at matched checkpoints.
By contrast, the commercial synthesis engine and the commercial P\&R 
engine used to \emph{produce} COMM solutions are left unnamed.
Aside from this tool anonymity, we emphasize that COMM is used strictly as 
a reference to contextualize trends observed with the open infrastructure,
rather than for any head-to-head tool comparison.

We evaluate three benchmark designs (\textit{aes}, \textit{ibex} and 
\textit{jpeg}) under five technology enablements: two 2D baselines
(\textit{ASAP7} and \textit{NanGate45}) and three 3D stacks
(\textit{ASAP7\_3D}, \textit{NanGate45\_3D}, and a heterogeneous stack denoted
as \textit{ASAP7+NanGate45\_3D})~\cite{asap7_repo, nangate45_pdk}. In our naming
convention, the first technology corresponds to the \emph{top} tier and the
second corresponds to the \emph{bottom} tier; for brevity, we also denote the
three 3D stacks as \textbf{7+7}, \textbf{45+45}, and \textbf{7+45}.

\begin{table}[h]
  \centering
  \caption{Hybrid Bonding Terminal (HBT) geometry settings.}
  \label{tab:hbt_settings}
  \resizebox{0.75\columnwidth}{!}{%
  \begin{tabular}{lrr}
    \toprule
    \textbf{Parameter} & \textbf{Value} & \textbf{Description} \\
    \midrule
    Width & $0.5\,\mu\mathrm{m}$ & Side length of the square cut \\
    Spacing & $0.5\,\mu\mathrm{m}$ & Minimum edge-to-edge distance \\
    Pitch & $1.0\,\mu\mathrm{m}$ & Effective center-to-center pitch \\
    Resistance & $0.02\,\Omega$ & Parasitic resistance per HBT \\
    \bottomrule
  \end{tabular}%
  }
  \vspace{-1em}
\end{table}

Table~\ref{tab:hbt_settings} lists the nominal geometry for Hybrid Bonding Terminals (HBTs)
used in our main experiments.
We model HBTs as square inter-tier vias with a $1.0\,\mu\mathrm{m}$ pitch.
Since this size is fixed, its impact varies across technology nodes.
In 45+45, the blockage caused by HBTs is manageable.
In 7+7, however, an HBT spans more routing tracks,
which increases congestion and makes pin access more difficult.
To illuminate sensitivities to scaling, Figure~\ref{fig:hb_pitch_analysis} shows 
results from varying HBT pitch while maintaining
$\text{WIDTH}=\text{SPACING}$ and $\text{PITCH}=\text{WIDTH}+\text{SPACING}$.

In all experiments, we target a fixed 60\% core utilization and derive the
die outline from the synthesized design size.\footnote{
In Table~\ref{tab:pin_3d_results}, \emph{Core} is the footprint area of a
single tier. We assume symmetric floorplans, so both tiers share the same
outline. \emph{StdCell} is the total standard-cell area summed over the top
and bottom tiers, including logic, buffers, and physical-only cells.
For the 2D results in Table~\ref{tab:2d_results}, \emph{Core} and \emph{StdCell}
follow standard single-die definitions.}
Both toolchains are evaluated at
the typical--typical (TT) corner. Parasitics are extracted using a unified 3D
technology LEF that captures the 3D stack definition. Power is estimated in
vectorless mode with a default switching activity, so the reported differences
primarily reflect physical implementation effects rather than stimulus
variations. For routing, we limit the detailed-routing engine to 20 iterations
to generate the final routed DEF.

We collect consistent metrics, including runtime, memory, routed wirelength (rWL), 
timing (WNS/TNS), and physical/timing violations (Design Rule Violations 
(DRVs) and Failing Endpoints (FEPs)).

\begin{table}[!htbp]
\centering
\caption{3D implementation results across stacks, designs, and flows.
Results are obtained using the Flexible tier strategy.}
\label{tab:pin_3d_results}
\resizebox{0.48\textwidth}{!}{%
\begin{tabular}{@{}c c c r r r r r r r r r r@{}}
\toprule
\multirow{2}{*}{\textbf{Enablement}} &
\multirow{2}{*}{\textbf{Design}} &
\multirow{2}{*}{\textbf{Flow}} &
\textbf{Clock} &
\multicolumn{2}{c}{\textbf{Area ($\mu\mathrm{m}^2$)}} &
\textbf{Power} &
\textbf{rWL} &
\multicolumn{2}{c}{\textbf{Timing (ns)}} &
\multicolumn{2}{c}{\textbf{Violations}} &
\textbf{HBT} \\
\cmidrule(lr){5-6} \cmidrule(lr){9-10} \cmidrule(lr){11-12}
& & &
\textbf{(ns)} &
\textbf{Core} &
\textbf{StdCell} &
\textbf{(mW)} &
\textbf{(mm)} &
\textbf{WNS} &
\textbf{TNS} &
\textbf{DRVs} &
\textbf{FEPs} &
\textbf{Cnt} \\
\midrule

\multirow{6}{*}{\textbf{45+45}} &
aes  & COMM & 0.820 & 13,342.6 & {16,446.5} & {20.34}  & {232.2} & {-0.016} & {-0.043}   & {3}  & {6}   & {904} \\
&     & ORD  & 0.820 & 15,262.0 & {18,100.8} & {46.34}  & {193.4} & {-0.064} & {-1.092}   & {8}  & {55}  & {650} \\
\cmidrule(lr){2-13}
& ibex & COMM & 2.200 & 18,295.3 & {22,244.2} & {11.35}  & {228.8} & {0.147}  & {0.000}    & {8}  & {0}   & {1,072} \\
&      & ORD  & 2.200 & {24,022.5} & {30,307.2} & {25.80}  & {291.6} & {-0.508} & {-255.542} & {1}  & {1,175} & {2,041} \\
\cmidrule(lr){2-13}
& jpeg & COMM & 1.200 & 60,049.5 & {68,321.0} & {75.05}  & {386.0} & {0.086}  & {0.000}    & {13} & {0}   & {1,492} \\
&      & ORD  & 1.200 & 86,059.5 & {98,157.5} & {122.10} & {614.3} & {0.008}  & {0.000}   & {2}  & {0}   & {2,880} \\
\midrule

\multirow{6}{*}{\textbf{7+7}} &
aes  & COMM & 0.380 & {892.8} & {1,260.2} & {7.20}  & {57.6}  & {-0.020} & {-0.305}   & {5,415}  & {64}   & {1,111} \\
&     & ORD  & 0.380 & {1,410.5} & {1,866.8} & {25.53}  & {70.8}  & {-0.087} & {-4.957}   & {16,779}  & {156}  & {2,261} \\
\cmidrule(lr){2-13}
& ibex & COMM & 1.000 & {1,340.4} & {1,650.1} & {5.71}  & {72.4}  & {-0.009} & {-0.064}   & {3,119}  & {27}  & {1,027} \\
&      & ORD  & 1.000 & {1,965.5} & {2,655.7} & {10.99} & {128.4} & {-0.449} & {-260.812} & {39,311} & {1,371} & {4,466} \\
\cmidrule(lr){2-13}
& jpeg & COMM & 0.680 & {3,532.6} & {4,064.9} & {26.20} & {103.6} & {-0.002} & {-0.002}   & {2,632}  & {1}    & {1,445} \\
&      & ORD  & 0.680 & {5,445.1} & {6,788.3} & {47.68} & {158.2} & {0.061}  & {0.000}    & {14,270} & {0}    & {3,328} \\
\midrule

\multirow{6}{*}{\textbf{7+45}} &
aes  & COMM & 0.820 & {1,958.5} & {2,311.8} & {5.96}  & {80.4}  & {-0.607} & {-60.910}   & {19}  & {227}   & {122} \\
&     & ORD  & 0.820 & {1,946.4} & {2,443.6} & {7.84}  & {64.7}  & {0.052}  & {0.000}     & {217}  & {0}    & {88} \\
\cmidrule(lr){2-13}
& ibex & COMM & 2.200 & {2,611.7} & {3,147.0} & {7.69}  & {78.8}  & {-1.264} & {-438.321}  & {124}    & {539}    & {257} \\
&      & ORD  & 2.200 & {4,501.7} & {4,823.4} & {5.23}  & {110.6} & {-0.043} & {-0.043}    & {1,351}  & {1}    & {671} \\
\cmidrule(lr){2-13}
& jpeg & COMM & 1.200 & {7,976.3} & {9,342.2} & {20.3}  & {133.9} & {0.006}  & {0.000}     & {11}     & {0}    & {491} \\
&      & ORD  & 1.200 & {11,927.9} & {13,820.0} & {29.92} & {236.4} & {0.241}  & {0.000}     & {3,686}  & {0}    & {1,946} \\
\bottomrule
\end{tabular}
}
\end{table}

Table~\ref{tab:pin_3d_results} and Table~\ref{tab:2d_results} respectively
summarize representative 3D and 2D results
under a common evaluation contract.\footnote{For the 7+45 case, \textbf{COMM} is evaluated by importing a saved design database at the reporting
checkpoint (instead of direct DEF readback) to avoid import-related errors. For \textbf{ORD}, metrics are
extracted by reading DEF, Verilog, and SDC at the matched checkpoint. Numbers are shown only to demonstrate
end-to-end flow completeness and are not intended for any direct comparison.}
The purpose of these tables is to demonstrate the \emph{flexibility} of the 
ORFS- and RosettaStone~2.0-based implementation and the Pin-3D enablements: 
the infrastructure enables controlled studies across stacks,
benchmarks, and toolchains with consistent checkpoints and metric semantics.
Notably, the 3D results (Table~\ref{tab:pin_3d_results}) are produced using 
ORFS-Research to exercise the Pin-3D enablements, whereas the 2D baselines
(Table~\ref{tab:2d_results}) use the default 2D ORFS flow \cite{ORFS}.
This split is intentional: it highlights that RosettaStone~2.0 can accommodate both 
stable mainline flows and research-enablements within a unified benchmarking 
and reporting framework.

Table~\ref{tab:pin_3d_results} reports relatively high HBT counts, 
especially for the 7+7 stack. A primary contributor to this is cross-tier 
buffer insertion, which can typically add up to two extra HBTs (and increase
routing complexity) when both incident nets become cross-tier.
Prior work~\cite{pentapati_pin-3d_2024} shows that strictly forbidding
cross-tier buffering can reduce HBT usage, but may also
degrade overall QoR.
In our reference flow, we keep this optimization enabled for robustness and
report the resulting HBT counts for transparency.\footnote{We have separately 
examined \textbf{ORD} flow
outcomes when the \SI{0.380}{ns} target clock period for \emph{aes} in 
Table~\ref{tab:pin_3d_results} is relaxed to
\SI{5}{ns}. TritonPart reports a cutsize of 147 (IO excluded). With the
\SI{5}{ns} clock period, the final 
layout contains 784 HBTs (reduced from 2,261) and 651 cross-tier 
nets (229 IO, 422 internal); the final DRC count is 96 (reduced 
from 16,779). The remaining cross-tier connectivity and the gap between 
HBT count and physical cross-tier net count mainly come from cross-tier 
buffer insertion, CTS clock buffering, and subsequent optimization.}

We also clarify the main DRC contributors in the 7+7 stack.
Because our HBT geometry is large relative to the 7+7 routing
pitch, the DRCs of COMM flow are dominated by HBT-induced
\texttt{SHORT}/\texttt{CUTSPACING} violations originating
from the relatively oversized HBT features under 7+7 rules.
To keep the evaluation contract consistent across all settings, we cap detailed
routing to only 20 iterations in all flows, which can leave residual violations.

\begin{table}[!htbp]
\centering
\caption{2D implementation results using the default 2D ORFS flow and an unnamed 2D 
commercial flow.}
\label{tab:2d_results}
\resizebox{0.48\textwidth}{!}{%
\begin{tabular}{@{}c c c r r r r r r r r r@{}}
\toprule
\multirow{2}{*}{\textbf{Enablement}} & \multirow{2}{*}{\textbf{Design}} & \multirow{2}{*}{\textbf{Flow}} & \textbf{Clock} & \multicolumn{2}{c}{\textbf{Area ($\mu\mathrm{m}^2$)}} & \textbf{Power} & \textbf{rWL} & \multicolumn{2}{c}{\textbf{Timing (ns)}} & \multicolumn{2}{c}{\textbf{Violations}} \\
\cmidrule(lr){5-6} \cmidrule(lr){9-10} \cmidrule(lr){11-12}
& & & \textbf{(ns)} & \textbf{Core} & \textbf{StdCell} & \textbf{(mW)} & \textbf{(mm)} & \textbf{WNS} & \textbf{TNS} & \textbf{DRVs} & \textbf{FEPs} \\
\midrule
\multirow{6}{*}{\textbf{NanGate45}} & aes & COMM & 0.820 & 28,870.8  & 16,157.1 & {21.6} & {249.3} & -0.013 & -0.074  & 0 & 17  \\
&  & ORFS  & 0.820 & {28,376.9}  & {20,142.6} & {48.9} & {212.9} & {-0.037} & {-0.681}  & {0} & {47} \\
\cmidrule(lr){2-12}
& ibex & COMM & 2.200 & 36,842.9  & 22,094.0 & {12.1} & {229.0} & 0.019  & 0.000   & 0 & 0   \\
&  & ORFS  & 2.200 & {47,332.0}  & {29,529.5} & {20.6} & {257.7} & {-0.094} & {-5.401} & {0} & {247} \\
\cmidrule(lr){2-12}
& jpeg & COMM & 1.200 & 114,557.4 & 68,179.0 & {78.4} & {397.4} & 0.011  & 0.000   & 0 & 0   \\
&  & ORFS  & 1.200 & {147,735.9} & {93,869.5} & {143.6} & {552.5} & {-0.189} & {-45.772}  & {0} & {702} \\
\midrule
\multirow{6}{*}{\textbf{ASAP7}} & aes & COMM & 0.380 & 1,691.7 & 1,229.6 & {6.6}  & {52.6}  & {-0.005} & {-0.020} & {0}      & {9}  \\
&  & ORFS  & 0.380 & {2,786.2} & {1,950.5} & {30.9} & {62.3} & {-0.041} & {-2.498} & {161} & {148} \\
\cmidrule(lr){2-12}
& ibex & COMM & 1.000 & 2,515.7 & 1,579.2 & {5.2}  & {65.6}  & {-0.016} & {-0.891} & {0}      & {152} \\
&  & ORFS  & 1.000 & {3,512.3} & {2,445.3} & {10.9} & {78.7} & {-0.078} & {-38.599} & {564} & {977} \\
\cmidrule(lr){2-12}
& jpeg & COMM & 0.680 & 6,505.7 & 4,016.0 & {24.7} & {99.6}  & {0.001}  & {0.000}  & {3}      & {0}   \\
&  & ORFS  & 0.680 & {10,207.9} & {6,410.3} & {46.4} & {158.0} & {0.008} & {0.000} & {1,249} & {0}   \\
\bottomrule
\end{tabular}
}
\end{table}

\medskip
\noindent\textbf{Impact of tier strategy.}
Table~\ref{tab:tier_strategy_comparison} gives example comparisons between the
\textbf{Restricted} and \textbf{Flexible} tier strategies across stacks and designs.
The observed effects vary by design and stack, and changes in timing, area/power, 
and HBT usage do not always move in the same direction. 

\begin{table}[h]
  \centering
  \caption{Effect of tier strategy on flow outcomes.}
  \label{tab:tier_strategy_comparison}
  \resizebox{0.48\textwidth}{!}{%
  \begin{tabular}{@{} l l l
        S[table-format=5.1]                                     
        S[table-format=3.1,round-mode=places,round-precision=1] 
        S[table-format=3.1,round-mode=places,round-precision=1] 
        S[table-format=+1.3]                                    
        S[table-format=-3.3]                                    
        S[table-format=4.0] @{}}
  \toprule
   \multirow{2}{*}{\textbf{Enablement}} & \multirow{2}{*}{\textbf{Design}} & \multirow{2}{*}{\textbf{Strategy}} &
  {\textbf{StdCell Area}} & {\textbf{Power}} & {\textbf{rWL}} &
  {\textbf{WNS}} & {\textbf{TNS}} & {\textbf{HBT}} \\
    & & &
    {($\mu\mathrm{m}^2$)} & {(mW)} & {(mm)} &
    {(ns)} & {(ns)} & {Count} \\
  \midrule

  \multirow{6}{*}{\shortstack[l]{\textbf{45+45}\\\textbf{ORD}}}
  & \multirow{2}{*}{aes}
    & Restricted & 18109.5 & 46.3929  & 197.6690 & -0.132 & -2.110   & 749  \\
  & & Flexible  & 18100.8 & 46.3385  & 193.3579 & -0.064 & -1.092   & 650  \\
  \cmidrule(lr){2-9}

  & \multirow{2}{*}{ibex}
    & Restricted & 30423.2 & 26.3642  & 311.3324 & -1.175 & -755.922 & 2432 \\
  & & Flexible  & 30307.2 & 25.8001  & 291.6397 & -0.508 & -255.542 & 2041 \\
  \cmidrule(lr){2-9}

  & \multirow{2}{*}{jpeg}
    & Restricted & 98168.6 & 122.6422 & 629.5467 & -0.159 & -1.106   & 3055 \\
  & & Flexible  & 98157.5 & 122.1016 & 614.3220 & +0.008 & 0.000    & 2880 \\
  \midrule

  \multirow{6}{*}{\shortstack[l]{\textbf{7+7}\\\textbf{COMM}}}
  & \multirow{2}{*}{aes}
    & Restricted & 1420.2 & 7.5660  & 58.3524  & -0.014 & -0.669   & 1097 \\
  & & Flexible  & 1260.2 & 7.1985  & 57.5683  & -0.020 & -0.305   & 1111 \\
  \cmidrule(lr){2-9}

  & \multirow{2}{*}{ibex}
    & Restricted & 1659.8 & 5.7297  & 71.1923  & -0.022 & -4.206   & 993  \\
  & & Flexible  & 1650.1 & 5.7077  & 72.4015  & -0.009 & -0.064   & 1027 \\
  \cmidrule(lr){2-9}

  & \multirow{2}{*}{jpeg}
    & Restricted & 4084.9 & 26.5308 & 102.9201 & +0.003 & 0.000    & 1364 \\
  & & Flexible  & 4064.9 & 26.1958 & 103.5957 & -0.002 & -0.002   & 1445 \\
  \midrule

  \multirow{6}{*}{\shortstack[l]{\textbf{7+45}\\\textbf{COMM}}}
  & \multirow{2}{*}{aes}
    & Restricted & 2295.5 & 4.4141 & 80.3633 & -0.004 & -0.005   & 43  \\
  & & Flexible  & 2311.8 & 5.9638 & 80.4241 & -0.607 & -60.910  & 122 \\
  \cmidrule(lr){2-9}

  & \multirow{2}{*}{ibex}
    & Restricted & 3107.6 & 2.6241 & 77.6072 & +0.110 & 0.000    & 260 \\
  & & Flexible  & 3147.0 & 7.6947 & 78.7585 & -1.264 & -438.321 & 257 \\
  \cmidrule(lr){2-9}

  & \multirow{2}{*}{jpeg}
    & Restricted & 9343.8 & 20.3147 & 137.4607 & +0.010 & 0.000   & 501 \\
  & & Flexible  & 9342.2 & 20.2770 & 133.9381 & +0.006 & 0.000   & 491 \\
  \bottomrule
  \end{tabular}%
  }
  \vspace{-1em}
\end{table}

\medskip
\noindent\textbf{Analysis of QoR gap contributors.}
To explore differences between ORD and COMM flow outcomes, we run \textit{aes} in the
45+45 enablement
under six mixed toolchain configurations:
(1) \textbf{COMM}: commercial synthesis + commercial backend;
(2) \textbf{Mixed1}: commercial synthesis + OpenROAD placement + commercial CTS/routing;
(3) \textbf{Mixed2}: commercial synthesis + OpenROAD backend;
(4) \textbf{Mixed3}: Yosys synthesis + commercial backend;
(5) \textbf{Mixed4}: Yosys synthesis + commercial placement + OpenROAD CTS/routing;
(6) \textbf{ORD}: Yosys synthesis + OpenROAD backend.

As shown in Table~\ref{tab:synthesis_impact}, \textbf{Mixed1} achieves outcomes similar
to those of \textbf{COMM}, 
suggesting that OpenROAD placement is not a primary bottleneck.
In contrast, replacing commercial synthesis with Yosys (\textbf{Mixed3}) 
noticeably increases power; this points to synthesis as an area for
improvement. 
Degradation also occurs in the backend:  switching to OpenROAD 
CTS/routing (\textbf{Mixed1}$\rightarrow$\textbf{Mixed2},
\textbf{Mixed3}$\rightarrow$\textbf{Mixed4}) leads to a modest increase 
in dynamic power and implementation violations.

\begin{table}[h]
  \centering
  \caption{QoR comparison across mixed toolchain configurations (\textit{aes}, 45+45).}
  \label{tab:synthesis_impact}
  \resizebox{0.45\textwidth}{!}{%
  \begin{tabular}{l
                  S[table-format=5.1]
                  S[table-format=3.4]
                  S[table-format=3.1]
                  S[table-format=-1.3]
                  S[table-format=5.0]
                  S[table-format=3.0]
                  S[table-format=4.0]}
    \toprule
    \textbf{Config} &
    {\textbf{StdCell Area}} &
    {\textbf{Power}} &
    {\textbf{rWL}} &
    {\textbf{WNS}} &
    {\textbf{DRVs}} &
    {\textbf{FEPs}} &
    {\textbf{HBT}} \\
    &
    {($\mu\mathrm{m}^2$)} &
    {(mW)} &
    {(mm)} &
    {(ns)} &
    & & \\
    \midrule
    COMM   & 16446.5 & 20.3358 & 232.2 & -0.016 & {3}   & {6}  & {904} \\
    Mixed1 & {16510.1} & {22.5075} & {237.5} & {-0.040} & {0}   & {54} & {1031} \\
    Mixed2 & {16193.3} & {24.5887} & {209.8} & {-0.135} & {153} & {144} & {522} \\
    Mixed3 & {17604.7} & {38.7790} & {195.2} & {0.012}  & {0}   & {0}  & {982} \\
    Mixed4 & {17664.3} & {45.2540} & {189.7} & {-0.094} & {0}   & {87} & {915} \\
    ORD    & {18100.8} & {46.3385} & {193.4} & {-0.064} & {8}   & {55} & {650} \\
    \bottomrule
  \end{tabular}%
  }
  \vspace{-1em}
\end{table}

\medskip
\noindent\textbf{Runtime analysis.}
We also examine stage-wise runtimes of the Pin-3D flow under 
OpenROAD-Research and the commercial reference flow, across
toolchains and technologies.
Figure~\ref{fig:jpeg_runtime_analysis} shows the runtime breakdown for the
\textit{jpeg} design across three 3D configurations.
Routing dominates the total runtime in all cases.
While OpenROAD-Research shows efficiency in pre-routing steps such as PDN,
placement, and legalization, its routing is much slower, leading to longer 
overall runtimes. This appears to be mainly driven by the high number of 
metal layers and persistent DRVs, and presents a clear opportunity for
tool improvement.

\medskip
\noindent\textbf{Hybrid Bonding Terminal pitch sweep.}
Figure~\ref{fig:hb_pitch_analysis} sweeps the hybrid bonding terminal (HBT) pitch 
for the testcase \textit{aes}, 7+7. 
(As noted earlier, the HBT geometry is defined such that $\text{WIDTH}=\text{SPACING}$ 
and the pitch equals $\text{WIDTH}+\text{SPACING}$.)
As HBT pitch shrinks beyond a certain point, the total DRV count drops 
in both ORD and COMM. Notably, when HBT feature size is reduced to be comparable 
to a typical upper-metal via (\texttt{V6\_m}-level pitch size), the DRV count 
for \textit{aes}, 7+7 drops to a near-zero level (51 and 88, respectively) in the 
ORD and COMM flows. 

Figure~\ref{fig:hb_pitch_drc} breaks down DRVs by layer. At larger HBT pitch,
violations concentrate on the HBT layer and its adjacent metal layers \texttt{M6\_m} and \texttt{M7}. 
As HBT pitch decreases, the HBT layer-related portion shrinks and the HBT-related \texttt{SHORT} and
\texttt{CUTSPACING} types largely vanish, consistent with the drop in total DRVs.%

\medskip
\noindent\textbf{Clock-period sweep.}
Figure~\ref{fig:clock_period_analysis} studies the impact of clock constraints on
implementation closure for \textit{aes}, 45+45. We sweep the target
clock period in \SI{0.05}{ns} steps. For each target clock period, we re-run logic
synthesis and then keep a fixed 60\% target utilization for floorplanning, so that
the observed QoR trends primarily reflect timing pressure induced by the
constraint.

To reduce run-to-run noise, we apply a simple local denoising procedure.
For each target clock period $T$, we also run the two neighboring settings
at $T-\SI{0.01}{ns}$ and $T+\SI{0.01}{ns}$, and report each metric as the
average over $\{T-\SI{0.01}{ns},\,T,\,T+\SI{0.01}{ns}\}$ outcomes.

Higher target frequency typically increases both timing and routability pressure.
As the clock period decreases, meeting timing requires more aggressive optimization,
typically increasing buffering and total cell count and raising routing demand.
Accordingly, rWL and total power tend to increase under tighter clocks,
and endpoint repair becomes more difficult.

In the 3D setting, tighter clocks can also increase cross-tier connectivity.
More cross-tier connections imply more HBTs, which add blockage
and routing constraints in the unified-technology abstraction.
As a result, DRVs can rise from near-zero levels as the clock period shrinks, indicating
increased routing difficulty around HBT features under higher timing pressure.

\section{Roadmap to RosettaStone~2.0}
\label{sec:rs2_roadmap}

RosettaStone~2.0 is designed as a maintained component of the
OpenROAD-Research project, enabling seamless integration between legacy
academic benchmarks, modern physical design data models, and end-to-end
evaluation flows. 
In contrast to RosettaStone~1.0~\cite{kahng_rosettastone_2022}, which relied 
on external and loosely coupled conversion utilities, 
RosettaStone~2.0 is tightly coupled with
OpenROAD-Research and ORFS-Research to ensure long-term sustainability.

Beyond simple format conversion, RosettaStone~2.0 introduces benchmark
normalization and repair mechanisms to sanitize legacy inputs. 
Notably, it automatically filters or fixes common inconsistencies found 
in academic benchmarks. Examples of this include removing ill-formed nets 
(nets with all input or output pin) and splitting overly-large standard-cell 
instances into multiple smaller ones. Such repairs preserve the original 
design’s intent while making the benchmarks valid for modern physical design 
tools, thereby ensuring physical realizability and fair
evaluation of results. The framework also supports a wide range of benchmark
formats, including traditional Bookshelf contest benchmarks from past 
academic competitions, as well as simplified or incomplete “fake” LEF/DEF 
design representations used in placement research. This broad compatibility 
means that RosettaStone~2.0 can ingest legacy benchmarks and bring
them into a consistent, up-to-date RTL-to-GDS flow.

To address the limited availability of real-world designs for stress-testing
algorithms, RosettaStone~2.0 also integrates with 
ArtNet, a hierarchical
clustering-based synthetic netlist generator. ArtNet can produce scalable
artificial circuits that mimic realistic design characteristics under
user-controlled parameters (size, hierarchy, etc.). This allows researchers to
generate new testcases on demand, complementing the fixed set of public
benchmarks with structurally representative synthetic designs. 
All such cases are handled through the same unified
RosettaStone interface and benefit from the framework’s normalization and
logging features. Together, these enhancements make RosettaStone~2.0 a robust
and extensible benchmarking platform that lowers entry barriers and enables
fair comparisons across both 2D and emerging 3D physical design flows,
including the Pin-3D flow.


We now summarize the supported input classes, translation paths, 
and evaluation methodology that together instantiate the RosettaStone~2.0 flow.

\smallskip
\noindent\textbf{A. Input classes.}
RosettaStone~2.0 supports multiple classes of academic benchmarks: traditional
Bookshelf-based contest benchmarks, simplified or ``fake'' LEF/DEF
representations common in placement research, and synthetic designs. To scale
the number of evaluation instances beyond finite public benchmarks, 
ArtNet~\cite{kang_artnet_2025,artnet_repo} can be used to generate structurally
representative, configurable, and reproducible netlists. Each synthetic design
is packaged with standardized enablements, ensuring that synthetic and real
designs share the same flow stages and measurement methodology.

\medskip
\noindent\textbf{B. Translation paths.}
For Bookshelf benchmarks, RosettaStone~2.0 directly translates the input into
the OpenROAD internal database format (.odb) via OpenDB~\cite{opendb} APIs, preserving
placement and netlist data while resolving mismatches in scale, grid alignment,
and technology abstraction. For benchmarks provided in incomplete or ``fake''
LEF/DEF formats, which often lack sufficient PDK information, RosettaStone~2.0
follows a complementary path: it first reconstructs an equivalent Bookshelf
representation, then remaps this form
to target PDKs to generate technology-consistent LEF/DEF files compliant with
modern design rules. This ensures that even academic benchmarks can be
evaluated in realistic, technology-aware settings.
 
\medskip
\noindent\textbf{C. Evaluation and metrics.}
Once the .odb is generated, OpenROAD-Research executes an end-to-end
RTL-to-GDS-style backend flow, while ORFS-Research provides structured flow
control. RosettaStone~2.0 aligns with the METRICS2.1 convention and
ORFS-Research’s structured logging infrastructure, enabling transparent
reporting of runtime, quality-of-results (QoR), and intermediate-stage 
metrics. By embedding this framework into the OpenROAD-Research 
continuous integration (CI) pipeline, we ensure that results remain 
reproducible and that the evaluation backplane evolves consistently 
with the underlying tools.

\section{Conclusion}
\label{sec:conclusion}

We contribute toward the goal of sustainable and transparent benchmarking 
for academic physical design research, by establishing RosettaStone~2.0 as 
a maintained backplane within the OpenROAD-Research, ORFS-Research 
ecosystem. The backplane co-versions artifacts with the toolchain, continuously 
validates regressions via CI, and standardizes evaluation through METRICS2.1-style
structured reporting with community-facing governance. As an end-to-end example, 
we release and validate a Pin-3D-style F2F hybrid-bonded 3D RTL-to-GDS reference 
flow with 3D enablements and stage-wise checkpoints. Finally, we outline a 
roadmap toward a complete RosettaStone~2.0 framework that systematically supports 
benchmark generation/expansion and unified 2D/3D flow validation.

\section*{Acknowledgment}
The work of Liwen Jiang, Zhiang Wang and Zhiyu Zheng is supported partly by AI for Science Program, Shanghai Municipal Commission of Economy and Informatization (2025-GZL-RGZN-BTBX-02038),
and partly by Fudan Kunpeng\&Ascend Center of Cultivation.
Experimental studies are performed at the Fudan Kunpeng\&Ascend Center of Cultivation, Fudan University. 
We thank Sayak Kundu and Matt Liberty for helpful discussions of the Pin-3D enablement and flow.

\clearpage


\footnotesize
\balance

\end{document}